Title:
Metabolomic measures of altered energy metabolism mediate the relationship of inflammatory miRNAs to motor control in collegiate football athletes

Running Title:
Integrating miRNA, metabolomics, and behavior

Nicole L. Vike[1,9], Sumra Bari[1,9], Khrystyna Stetsiv[1], Linda Papa[2,10], Eric A. Nauman[3,4,5,10], Thomas M. Talavage[3,6,10], Semyon Slobounov[7,10*], Hans C. Breiter[1,8,9*]

[1]Department of Psychiatry and Behavioral Sciences, Feinberg School of Medicine, Northwestern University, Chicago, IL, USA
[2] Department of Emergency Medicine, Orlando Regional Medical Center, Orlando, FL, USA
[3]Weldon School of Biomedical Engineering, Purdue University, West Lafayette, IN, USA
[4]School of Mechanical Engineering, Purdue University, West Lafayette, IN, USA
[5]Department of Basic Medical Sciences, Purdue University, West Lafayette, IN, USA
[6]School of Electrical and Computer Engineering, Purdue University, West Lafayette, IN, USA
[7]Department of Kinesiology, Pennsylvania State University, University Park, PA, USA;
[8]Laboratory of Neuroimaging and Genetics, Department of Psychiatry, Massachusetts General Hospital and Harvard School of Medicine, Boston, MA, USA

[9,10] indicate co-equal authorship

* Corresponding Authors:
For project design, management and data collection: Semyon Slobounov (sms18@psu.edu)
For hypotheses and conceptual framework, data analysis and paper development: Hans Breiter (h-breiter@northwestern.edu)

ABSTRACT

Recent research has shown there can be detrimental neurological effects of short- and long-term exposure to contact sports. In the present study, metabolomic profiling was combined with inflammatory miRNA quantification, computational behavior with virtual reality (VR) testing of motor control, and head collision event monitoring to explore *trans*-omic and collision effects on human behavior across a season of players on a collegiate American football team. We integrated permutation-based statistics with mediation analyses to test complex, directional relationships between miRNAs, metabolites, and VR task performance. Fourteen significant mediations (metabolite = mediator; miRNA = independent variable; VR score = dependent variable) were discovered at preseason (N=6) and across season (N=8) with Sobel *p*-values $\leq 0.05$ and with total effects at or exceeding 50%. The majority of mediation findings involved long to medium chain fatty acids (2-HG, 8-HOA, UND, sebacate, suberate, and heptanoate). In parallel, TCA metabolites were found to be significantly decreased at postseason relative to preseason. HAEs were associated with metabolomic measures and miRNA levels across-season. Together, these observations suggest a state of chronic HAE-induced neuroinflammation (as evidence by elevated miRNAs) and mitochondrial dysfunction (as observed by abnormal FAs and TCA metabolites) that together produce subtle changes in neurological function (as observed by impaired motor control behavior). These findings point to a shift in mitochondrial metabolism, away from mitochondria function, consistent with other illnesses classified as mitochondrial disorders, suggesting a plausible mechanism underlying HAEs in contact sports and potential avenue for treatment intervention.

INTRODUCTION

Sports-related concussion remains one of the most common causes of traumatic brain injuries (TBIs), with approximately 1.6-3.8 million occurrences annually [1, 2]. Despite the high incidence of concussive injuries in contact sports, the molecular mechanisms of these injuries are not well understood, making development of targeted pharmaceutical interventions challenging. Additionally, subconcussive injuries (i.e., head impacts yielding minimal diagnosable cognitive or behavioral deficits) have also been linked with significant neurophysiological and biochemical changes in youth, collegiate, and professional football athletes [3–7]. Due to the lack of observable subconcussive-related symptoms, research has sought to identify biomarkers based on advanced imaging and omics (e.g., transcriptomics, proteomics) [7–15].

Previous research by Papa and colleagues, and Bhomia and colleagues, elucidated a panel of potential miRNA biomarkers of football-related injury (miR-20a, miR-505, miR-92a, miR-195, miR-93p, miR-30d, miR-486, miR-3623p, and miR-151-5p) [16, 17]. This set of miRNAs was not only elevated at the end of the football season but was also found to be elevated prior to the season when compared to non-athletes reporting to the emergency room with potential concussion [16]. Findings from these studies support an inflammatory hypothesis, although we still know little about the molecular targets of these miRNAs. Inflammation is energy intensive, suggesting there may be alterations in metabolism and energy-related metabolites concurrent with miRNA increases. Indeed, MRS studies support this hypothesis [7, 18–21]. Furthermore, metabolomics appear to be important in the context of the diffuse axonal injury (DAI) model of mild TBI (mTBI), which models how shock-waves affect axons and their structure, which would suggest there may be significant effects on a broad array of fatty acids (FAs) involved with axonal function.

Metabolomics allows investigation of metabolic fluctuations following an experimental or observational perturbation, such as a season of contact sports participation. A number of molecular pathways have been studied in animal models and have been observed to be altered by head impacts [22–29]. Across animal and human studies of head impacts, mitochondrial dysfunction has been consistently observed [30–36]. Dubbed the powerhouse of the cell, mitochondria are critically important for energy production and cellular respiration [37, 38]. Damage to these organelles can thus result in serious cellular, and potentially systems-level, dysfunction.

One major role of the mitochondria is the oxidation of fatty acids into functional metabolites that can then fuel the TCA cycle and downstream energy production [39–41]. Long and medium chain fatty acids are first processed in peroxisomes, small organelles that oxidize fatty acids via alpha, omega, and beta oxidation [42–44]. Omega oxidation replaces the methyl terminus with a hydroxy group (now referred to as a monohydroxy fatty acid) which is then further oxidized into a carboxy group (dicarboxylic fatty acid). Once in dicarboxylic form, the fatty acid can then be further processed via beta oxidation in both the peroxisome and mitochondria [45, 46]. Depending on the length and processing location of the fatty acid, different sets of enzymes are utilized to metabolize it into a smaller, functional metabolite – acetyl-CoA. Acetyl-CoA then enters the TCA cycle to produce energy (GTP) and energy-storing (NADH and $FADH_2$) metabolites [47]. NADH and $FADH_2$ are then used in the electron transport chain to ultimately produce large amounts of ATP – the most fundamental source of energy. However, if mitochondrial damage occurs from repeated head acceleration events (HAEs) – either associated with direct impacts or whiplash motions resulting from collisions elsewhere on the body – metabolites involved in fatty acid oxidation (e.g. sebacate) and the TCA cycle (e.g. citrate) may be altered. Additionally, mitochondrial distress can result in oxidative stress and increased reactive oxygen species production [48–51]. Therefore, finding a set of metabolites to highlight potential mitochondrial dysfunction, while also showing relationships to neurological function/dysfunction and neuroinflammatory miRNAs, is a promising and novel biomarker.

While the discovery of blood biomarkers is the ultimate clinical goal, it is first necessary to establish if biomarkers have relationships with neurological function or dysfunction. Numerous studies have attempted to uncover behavioral changes in contact sport athletes, but many tests are insensitive to subtle changes in behavior [52–56]. Alexander Luria observed a set of behavioral abnormalities in individuals with mTBI that can imply a non-obvious alteration in biology [57–60]. These observations led to development of computational behavior tasks using virtual reality technology that can quantify abnormalities in motor control and are sensitive to motor control changes following TBI that may be missed in clinical examination [61–65].

This study specifically evaluated the hypothesis that an integrated system of omic measures, from the transcriptome (miRNA) and metabolome (individualized biochemistry), would form three-way associations with computational behavior measures prior to the season (preseason) or across the season (i.e., pre- vs. postseason) for a collegiate football team. We predicted these *trans*-omic and behavioral measures would show mediation relationships that were consistent (i.e., highly specific) in terms of the category of independent variable (IV), mediator variable (M), and dependent variable (DV), and that computational behavior measures would always be the DV. Given the large number of associations needing testing for this mediation analysis, we integrated permutation-based statistics with mediation analyses in a novel framework to address the issue of multiple comparisons. For behavioral tasks, we focused on motor control and spatial memory that were described by Luria as being abnormal without obvious head trauma [58]. For miRNA, we focused on a panel of miRNAs that had previously been shown to be abnormal relative to emergency room controls in the pre-season, and across-season for football players, and that were known to be involved in the control of inflammation [16]. To restrict the number of biochemical compounds identified with metabolomics, we focused only on metabolomic measures that were significantly different after correction for multiple comparisons across the season, and were (i) fatty acids and/or compounds involved with energy metabolism, (ii) compounds involved with stress/inflammatory responses, or (iii) exogenous compounds related to consumption. The focus of this *trans*-omics study was to determine if any metabolic pathway and its compounds was consistently observed in the majority of mediation relationships to be involved with abnormalities in inflammatory miRNAs and Luria-based behaviors. If such a pathway were found, it would present a potential unifying construct for head impact/concussion research and clinical intervention.

Following this approach, we identified a consistent abnormality in fatty acid compounds related to mitochondrial β-oxidation and downstream abnormalities in TCA metabolites. Alterations in four of the six metabolomic compounds were consistent with a dysfunction in β-oxidation and showed mediation relationships linking abnormalities in inflammatory miRNAs to abnormalities in Luria behaviors. A fifth compound observed in these mediation relationships (i.e., 2-hydroxyglutarate (2-HG)) was a known oncometabolite that has been implicated in altering the balance of oxidative metabolism with mitochondria in cancer cells. The sixth compound (i.e., adenosine) represents a fundamental ATP component. The majority of these metabolomic compounds were associated with head acceleration events (HAEs). Altogether, these *trans*-omic findings connecting two types of omic measures with quantitative variables from a well-controlled and ethologically grounded behavior, supports a hypothesis that repetitive HAEs produce a syndrome of abnormal mitochondrial metabolism.

MATERIALS/METHODS

Subjects and data collection

Twenty-four male collegiate American football players (mean age = $21 \pm 2$ years) were recruited for this study. Written informed consent was obtained from each subject in accordance with the Penn State University Institutional Review Board. Demographic information was obtained from each subject and confirmed by a team physician: age, years of play experience (YoE), player position, and history of diagnosed concussion (HoC; reported as the number of previous concussions). Blood samples were taken prior to any contact practices (preseason) and within one week of the last game (postseason). None of the subjects had a diagnosed concussion in the nine months preceding preseason data collection. Blood samples were prepared and sent out for miRNA quantification and metabolomic analysis. Due to missing miRNA data, the number of subjects decreased from 24 to between 21-23 depending on the miRNA of interest. Concurrent with blood collection, players also underwent virtual reality (VR) testing.

Serum extraction

Five mL of blood were collected from each subject at pre- and postseason. Samples were placed in a serum separator tube, allowed to clot at room temperature, and then centrifuged. Serum was extracted from each tube and pipetted into bar-coded aliquot tubes. Serum samples were stored at -70°C until they were transported to 1) a central laboratory for blinded miRNA batch analysis and 2) Metabolon (Morrisville, NC, USA) for blinded metabolite analysis.

Virtual reality testing

Virtual reality testing was conducted using a 3D TV system, head-mounted accelerometer, and interactive joystick device from Head Rehab LLC (Chicago, IL, USA). There were three separate VR tests: 1) balance (Bal), 2) whole-body reaction time (RT), and 3) spatial memory (SM). The scores from each test were combined to produce a comprehensive score (Comp). These tests were based off initial findings from Dr. Alexander Luria and have been validated to detect functional abnormalities following mild traumatic brain injury [57–59, 65, 66]. The Bal module tested subjects' abilities to maintain posture with a changing virtual environment. Subjects were instructed to hold a tandem Romberg position on a pressure plate for each trial [67]. The first trial collected baseline pressure data and the subsequent six trials tested performance. The virtual environment shifted in a different direction for each performance trial and deviances from baseline pressure points were reported. The RT module tested the time it took for subjects to adjust their posture in the direction of an altered virtual environment. Subjects stood shoulder-width apart with their hands on their hips and were instructed to move their hips in the direction of the virtual environment. Results were recorded using the pressure plate and reported as latency to shift the body. The SM module tested a subject's ability to recall a virtual pathway by first viewing and then recalling and navigating the pathway with the joystick; results were reported as the accuracy of correct responses relative to errors. More detailed descriptions of the tasks have been previously reported [68]. VR tests were conducted concurrent with blood sampling.

miRNA quantification

Serum samples collected at pre- and postseason were used to isolate and quantify levels of RNA. 100 $\mu$L of serum was aliquoted, and RNA was isolated using a serum/plasma isolation kit (Qiagen Inc., Venlo, Netherlands) as per the manufacturer's protocol. RNA was eluted in 20 $\mu L$ of DNAse/RNAse-free water and stored at -80°C until further use.

Droplet digital PCR (ddPCR; Bio-Rad Inc., Hercules, CA, USA) was used to quantify absolute levels of nine miRNA (miR-20a, miR-505, miR-3623p, miR-30d, miR-92a, miR-486, miR-195, miR-93p, miR-151-5p) [16]. Prior to ddPCR analysis, RNA was checked for quality using a bioanalyzer assay with a small RNA assay. After quality confirmation, 10 ng of RNA was reverse transcribed using specific miRNA TaqMan assays as per the manufacturer's protocol (Thermo Fisher Scientific Inc., Waltham, MA, USA). Protocol details can be found in [16]. The final PCR product was analyzed using a droplet reader (Bio-Rad Inc., Hercules, CA, USA). Total positive and negative droplets were quantified, and from this, the concentration of miRNA/$\mu L$ of the PCR reaction was reported. All reactions were performed in duplicate.

Metabolomic analysis

The remaining serum was sent to Metabolon (Morrisville, NC, USA) for metabolomic quantification. Upon arrival, samples were assigned a unique identifier via an automated laboratory system and stored at -80°C. Samples were prepared for subsequent analyses using an automated MicroLab STAR® system (Hamilton Company, Reno, NV, USA). Proteins were precipitated out of each sample using methanol and a shaker (Glen Mills GenoGrinder 2000), and then centrifuged. The resulting extract was then divided into five fractions for various analyses: 1) two fractions for analysis by two separate reverse phase (RP)/UPLC-MS/MS methods with positive ion mode electrospray ionization (ESI), 2) one for analysis by RP/UPLC-MS/MS with negative ion mode ESI, 3) one for analysis by HILIC/UPLC-MS/MS with negative ion mode ESI, and 4) one reserved for backup. To remove organic solvent, samples were briefly placed on a TurboVap® (Zymark); samples were then stored under nitrogen overnight prior to analyses.

Serum metabolites were quantified using Ultrahigh Performance Liquid Chromatography-Tandem Mass Spectroscopy (UPLC-MS/MS). All methods utilized a Waters ACQUITY UPLC and a Thermo Scientific Q-Exactive high resolution/accurate mass spectrometer interfaced with a heated electrospray ionization (HESI-II) source and Orbitrap mass analyzer operated at 35,000 mass resolution. The sample extract was dried and reconstituted in solvents compatible to each of the listed analyses. Each reconstitution solvent contained a series of standards at fixed concentrations to ensure injection and chromatographic consistency. One aliquot was analyzed using acidic positive ion conditions, chromatographically optimized for more hydrophilic compounds. In this method, the extract was gradient eluted from a C18 column (Waters UPLC BEH C18-2.1x100 mm, 1.7 μm) using water and methanol, containing 0.05% perfluoropentanoic acid (PFPA) and 0.1% formic acid (FA). Another aliquot was also analyzed using acidic positive ion conditions; however, it was chromatographically optimized for more hydrophobic compounds. In this method, the extract was gradient eluted from the same afore mentioned C18 column using methanol, acetonitrile, water, 0.05% PFPA and 0.01% FA and was operated at an overall higher organic content. Another aliquot was analyzed using basic negative ion optimized conditions using a separate dedicated C18 column. The basic extracts were gradient eluted from the column using methanol and water, however with 6.5mM Ammonium Bicarbonate at pH 8. The fourth aliquot was analyzed via negative ionization following elution from a HILIC column (Waters UPLC BEH Amide 2.1x150 mm, 1.7 μm) using a gradient consisting of water and acetonitrile with 10mM Ammonium Formate, pH 10.8. The MS analysis alternated between MS and data-dependent $MS^n$ scans using dynamic exclusion. The scan range varied slighted between methods but covered 70-1000 m/z.

Peak analysis was conducted using a bioinformatics system which consisted of four major components: 1) the Laboratory Information Management System (LIMS - a system used to automate sample accession and preparation, instrumental analysis and reporting, and data analysis), 2) the data extraction and peak-identification software, 3) data processing tools for quality control and compound identification, and 4) a collection of information interpretation and visualization tools.

Raw data were extracted, peak-identified, and QC processed using Metabolon's hardware and software. Compounds were identified by comparison to library entries of purified standards. Biochemical identifications were based on three criteria: 1) retention index (RI) within a narrow RI window of the proposed identification, 2) accurate mass match to the library (+/- 10 ppm), and 3) the MS/MS forward and reverse scores between the experimental data and authentic standards. The MS/MS scores were based on a comparison of the ions present in the experimental spectrum to the ions present in the library spectrum. While there may have be similarities between these molecules based on one of these factors, the use of all three data points was utilized to distinguish and differentiate more than 3,300 registered biochemicals.

Peaks were quantified using area-under-the-curve. A data normalization step was performed to correct variation resulting from instrument inter-day tuning differences (i.e. variation between pre and postseason analyses). Specifically, each compound was corrected in run-day blocks by registering the medians to equal one (1.00) and normalizing each data point proportionately. Data were then log-transformed. Of the 3,300+ potential biochemicals, 968 metabolites were analyzed at pre and postseason. Of these, 161 showed significant, FDR-corrected, increases or decreases between the pre and postseason (*q*-value < 0.05). Of those 161 metabolites, 40 were selected based on the following criteria: 1) they appeared in the random forest plot (20 of the 40 studied here) (Figure 1) and 2) they were in biochemical pathways which have been implicated to change following head impact events [69–73]. Based on these criteria, six macromolecule categories were defined: lipids (19/40), energy-related metabolites (5/40), xenobiotics (10/40), amino acids (3/40), carbohydrates (2/40), and nucleotides (1/40) (Table 1). All 40 metabolites listed in Table 1 were used in all subsequent analyses.

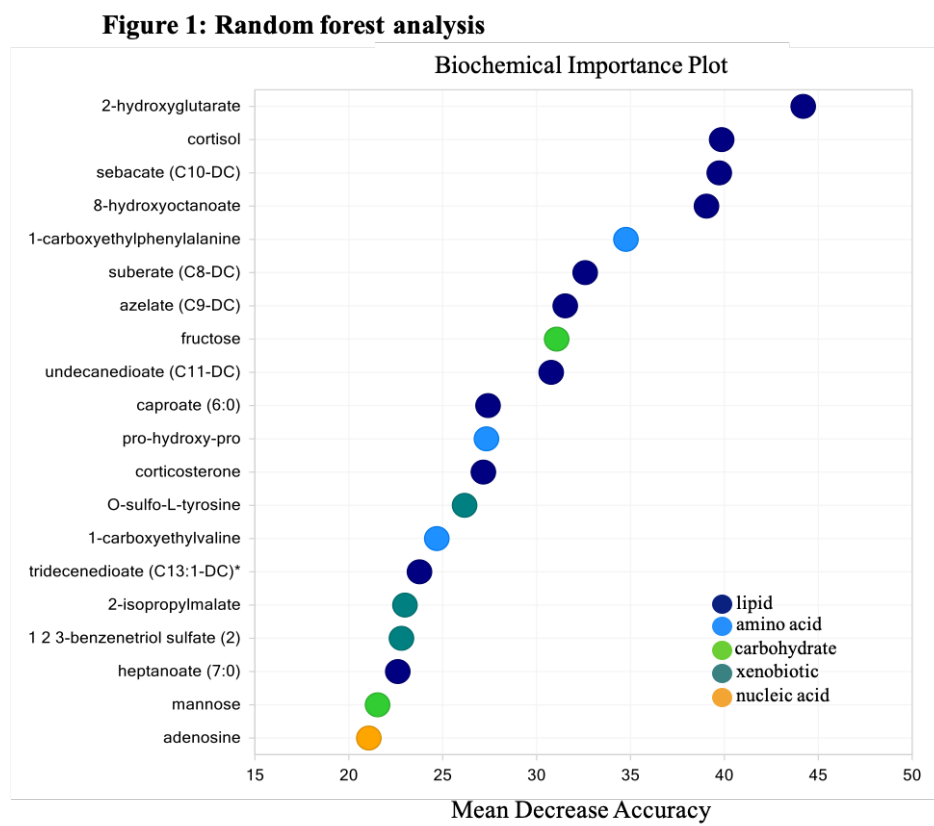

**Figure 1.** Random forest plot of metabolite importance. The mean decrease accuracy (MDA) is a measure of model performance when a particular metabolite was excluded from the analysis. As MDA increased, the more important that metabolite was to distinguish pre- versus post-season data. The metabolite with the highest MDA, and therefore, highest predictive power, was 2-hydroxyglutarate, a fatty acid. The majority of metabolites in the plot were lipids (55%).

## Table 1: Significant across-season metabolite changes

| Results from Signed-rank Test (Preseason vs. Postseason) | | | | | |
|---|---|---|---|---|---|
| Metabolite | Super Pathway | Sub Pathway | p-value | q-value | z-score |
| corticosterone* | lipid | corticosteroid | 0.0002 | 0.0028 | -3.74 |
| cortisol* | lipid | corticosteroid | 0.0001 | 0.0016 | -3.98 |
| cortisone | lipid | corticosteroid | 0.0447 | 0.0447 | -2.01 |
| cortoloneglucuronide (cg) | lipid | corticosteroid | 0.0040 | 0.0194 | 2.88 |
| stearidonate | lipid | long chain polyunsaturated fatty acid | 0.0035 | 0.0194 | -2.92 |
| linoleate3n6 | lipid | long chain polyunsaturated fatty acid | 0.0013 | 0.0091 | -3.22 |
| sebacate* | lipid | fatty acid, dicarboxylate | 0.0001 | 0.0016 | -3.95 |
| linoleate2n6 | lipid | long chain polyunsaturated fatty acid | 0.0097 | 0.0194 | -2.59 |
| dodecadienoate | lipid | fatty acid, dicarboxylate | 0.0068 | 0.0194 | -2.71 |
| azelate* | lipid | fatty acid, dicarboxylate | 0.0003 | 0.0036 | -3.62 |
| suberate* | lipid | fatty acid, dicarboxylate | 0.0002 | 0.0028 | -3.77 |
| 7-hydroxyoctanate | lipid | fatty acid, monohydroxy | 0.0074 | 0.0194 | -2.68 |
| 8-hydroxyoctanate* | lipid | fatty acid, monohydroxy | 0.0001 | 0.0016 | -3.86 |
| undecanedioate* | lipid | fatty acid, dicarboxylate | 0.0004 | 0.0040 | -3.53 |
| caproate* | lipid | medium chain fatty acid | 0.0004 | 0.0040 | 3.53 |
| heptanoate (7:0)* | lipid | medium chain fatty acid | 0.0008 | 0.0072 | 3.35 |
| tridecenedioate* | lipid | fatty acid, dicarboxylate | 0.0003 | 0.0036 | 3.65 |
| 2-hydroxyglutarate* | lipid | fatty acid, dicarboxylate | 0.0000 | 0.0000 | -4.14 |
| N-palmitoylserine | lipid | endocannabinoid | 0.0009 | 0.0072 | 3.34 |
| alpha-ketoglutarate | energy | TCA cycle | 0.0386 | 0.0386 | 2.07 |
| citrate | energy | TCA cycle | 0.0192 | 0.0384 | 2.34 |
| aconitate | energy | TCA cycle | 0.0068 | 0.0272 | 2.71 |
| fumarate | energy | TCA cycle | 0.0177 | 0.0384 | 2.37 |
| phosphate | energy | oxidative phosphorylation | 0.0051 | 0.0255 | -2.80 |
| paraxanthine | xenobiotic | xanthine | 0.0220 | 0.0440 | -2.29 |
| 1-methylurate | xenobiotic | xanthine | 0.0034 | 0.0272 | -2.93 |
| 1,3-dimethylurate | xenobiotic | xanthine | 0.0081 | 0.0440 | -2.65 |
| 1,7-dimethylurate | xenobiotic | xanthine | 0.0120 | 0.0440 | -2.51 |
| 1-methylxanthine | xenobiotic | xanthine | 0.0009 | 0.0081 | -3.32 |
| 5-acetylamino-6-amino-3-methyluracil (aam) | xenobiotic | xanthine | 0.0184 | 0.0440 | -2.36 |
| 5-acetylamino-6-formylamino-3-methyluracil (afm) | xenobiotic | xanthine | 0.0123 | 0.0440 | -2.50 |
| O-sulfo-L-tyrosine* | xenobiotic | chemical | 0.0001 | 0.0011 | -3.98 |
| 2-isopropylmalate* | xenobiotic | food component/plant | 0.0089 | 0.0440 | 2.62 |
| 1,2,3-benzenetriol sulfate (2)* | xenobiotic | chemical | 0.0005 | 0.0050 | -3.50 |
| 1-carboxyethylphenylalanine* | amino acid | phenylalanine metabolism | 0.0002 | 0.0002 | 3.74 |
| prolyl-hydroxy-proline* | amino acid | urea cycle; arginine and proline metabolism | 0.0000 | 0.0000 | -4.17 |
| 1-carboxyethylvaline* | amino acid | leucine, isoleucine and valine metabolism | 0.0002 | 0.0002 | 3.74 |
| fructose* | carbohydrate | fructose, mannose and galactose metabolism | 0.0001 | 0.0001 | 3.86 |
| mannose* | carbohydrate | fructose, mannose and galactose metabolism | 0.0001 | 0.0001 | -3.95 |
| adenosine* | nucleotide | polyamine metabolism | 0.0333 | 0.0333 | 2.13 |

**Table 1:** Metabolite changes across season. Metabolite levels between pre- and post-season were assessed using the Wilcoxon signed-rank test, a nonparametric *t*-test for paired samples. Metabolites were grouped by super pathway and FDR corrected by the number of variables in each group (e.g. 19 lipid metabolites). Metabolites with significant *q*-values were reported. Negative z-scores indicate an increase from pre- to post-season and positive z-scores indicate a decrease from pre- to post-season. * indicates metabolites that also appeared in the random forest plot.

Statistical analysis

All statistical analyses were performed in STATA (College Station, TX, USA) apart from the random forest analysis which was performed by Metabolon.

*Random Forest analysis*: Random forest is an unsupervised classification technique based on an ensemble of decision trees; here, it was used to assess the importance ranking of biochemicals (i.e., how well can a given metabolite be used to distinguish preseason from postseason) [74]. For a given decision tree, a random subset of the data, with identifying true class information, was selected to build the tree ("bootstrap sample" or "training set"), and then the remaining data, the "out-of-bag" (OOB) variables, were passed down the tree to obtain a class prediction for each sample. This process was repeated thousands of times to produce the forest. The final classification of each sample was determined by computing the class prediction frequency for the OOB variables over the whole forest. When the full forest was grown, the class predictions were compared to the true classes, generating the "OOB error rate" as a measure of prediction accuracy. Thus, the prediction accuracy was an unbiased estimate of how well a sample class was predicted in a new data set. To determine which biochemicals made the largest contribution to the classification, a variable importance measure, Mean Decrease Accuracy (MDA), was computed. MDA was determined by randomly permuting a variable, running the observed values through the trees, and then reassessing the prediction accuracy. If a variable was not important, the procedure produced little change in the accuracy of the class prediction (permuting random noise gave random noise). By contrast, if a variable was important to the classification, the prediction accuracy dropped; this was recorded as the MDA.

*Wilcoxon signed-rank test for across-season miRNA, metabolite, and VR score analysis*: Data were first tested for normality and equal variances using the Shapiro-Wilks test and Bartlett's test, respectively. Because 1) data were not normally distributed (Shapiro-Wilks test; $p \geq 0.05$) and/or data did not have equal variances (Bartlett's test; $p \geq 0.05$) and 2) data were paired by subject across-season, a Wilcoxon signed-rank test was utilized to assess across-season changes. Changes were analyzed at a significance level of 0.05. If *p*-value < 0.05, metabolites were then grouped based on super pathway and FDR-corrected using the Benjamini-Hochberg method for multiple comparisons (Table 1). Changes were considered significant if *q*-value ≤ 0.05.

*Linear regression analyses*: Linear regressions were conducted to assess significant interactions between VR scores and miRNAs, miRNAs and metabolites, and VR scores and metabolites, where VR score (i.e., behavior) was always the dependent variable (Y). When regressing miRNA and metabolite, metabolite was the designated as the dependent variable; this decision was based on previous findings where miRNA levels were elevated in this cohort at preseason when compared to controls [16]. After each initial regression analysis, Cook's distance was calculated to reveal outliers which drastically influenced the regression (i.e., large shift in β) [75, 76]. Outliers were removed if Cook's distance > 4/n and regressions were then re-run [77]. Regressions were considered significant if *p*-value < 0.05 and standardized beta coefficients (Std. β) and adjusted $R^2$ values ($R^2_{adj.}$) were also reported.

*Permutation-based mediation analysis*: Based on results from the linear regression analyses, data were prepared for mediation analysis by first assessing three-way associations. To be included in mediation analysis, all paths must have met significance (*p*-value < 0.05) to form a three-way association. Paths were as follows: A) miRNA (X) → metabolite (Y), B) metabolite (Y) → VR score (Z), and C) miRNA (X) → VR score (Z). Cook's outliers were removed across all regressions to achieve the same set of subjects, and regressions were re-run with common subjects. Mediation seeks to clarify the causal relationship between the independent variable (X) and dependent variable (Y) with the inclusion of a third variable mediator (M). The mediation model proposes that instead of a direct causal relationship between X and Y, the X influences M which then influences the Y. Beta coefficients (β) and standard error (se) terms from the following linear regression equations are used to calculate the Sobel $p$-value and mediation effect percentage ($T_{eff}$):

$$Step\ 1\ (Path\ A): M = \beta_0 + \beta_{1A}(X) + \epsilon_A$$

$$Step\ 2\ (Path\ B): Y = \beta_0 + \beta_{1B}(M) + \epsilon_B$$

$$Step\ 3\ (Path\ C,\ model\ 1): Y = \beta_0 + \beta_{1,1C}(X) + \epsilon_{1C}$$

$$Step\ 4\ (Path\ C,\ model\ 2): Y = \beta_0 + \beta_{1,2C}(X) + \beta_{2,2C}(M) + \epsilon_{2C}$$

Sobel's test was then used to test if $\beta_{1,2C}$ was significantly lower than $\beta_{1,1C}$ using the following equation:

$$(3)\ Sobel\ z-score = \frac{(\beta_{1,1C} - \beta_{1,2C})}{\sqrt{[(\beta_{2,2C})^2(\epsilon_{1A})^2] + [(\beta_{1A})^2(\epsilon_{2C})^2]}}$$

Using a standard 2-tail z-score table, the Sobel $p$-value was determined from Sobel z-score. Mediation effect percentage ($T_{eff}$) was calculated using the following equation:

$$(4)\ T_{eff} = 100 * \frac{(\beta_{1A} * \beta_{2,2C})}{(\beta_{1A} * \beta_{2,2C}) + [\beta_{1,1C} - (\beta_{1A} * \beta_{2,2C})]}$$

In order to control for the occurrence of false positives due to multiple hypotheses testing, permutation-based mediation analysis was utilized. Permutation tests re-sample observations from the original data multiple times to build empirical estimates of the null distribution for the test statistic being studied [78, 79]. Permutation-based tests are especially well-suited for studies with small sample sizes as they estimate the statistical significance directly from the data being analyzed rather than making assumptions about the underlying distribution. First, the test statistic is obtained from the original data set, then the data is randomly permuted multiple (S) times and the test statistic is computed on each permutated data set. The statistical significance is computed by counting (K) the number of times the statistic value obtained in the original data set was more extreme than the statistic value obtained from the permuted data sets, and dividing that value by the number of random permutations (K/S) [78].

Directed mediation analysis was performed with VR score always acting as the dependent variable (DV/Y); this *a priori* hypothesis has been observed in previous studies [80]. Secondly, for the hypothesis-directed mediation analyses, miRNA was always the independent variable (IV/X) and metabolite was always the mediator (M):

$$VR\ score = \beta_0 + \beta_{1,1C}(miRNA) + \epsilon_{1C}$$

$$VR\ score = \beta_0 + \beta_{1,2C}(miRNA) + \beta_{2,2C}(metabolite) + \epsilon_{2C}$$

The selection for IV was based off work by Papa et al. (2019) where miRNA levels were observed to be elevated in collegiate football athletes, compared to controls, at both pre- and postseason [16]. Therefore, metabolites, which can change dynamically, were hypothesized to mediate the relationship between elevated miRNAs and VR task performance. For control mediation analyses, VR score was DV, miRNA was M, and metabolite was IV:

$$VR\ score = \beta_0 + \beta_{1,1C}(metabolite) + \epsilon_{1C}$$

$$VR\ score = \beta_0 + \beta_{1,2C}(metabolite) + \beta_{2,2C}(miRNA) + \epsilon_{2C}$$

The measurement matrices $X_t, Y_t, M_t$ defined below schematize the variables used, miRNAs, VR scores, and metabolites respectively.

$$X_t = \begin{bmatrix} x_{t,1,1} & \cdots & x_{t,P,1} \\ \vdots & \cdots & \vdots \\ x_{t,1,N} & \cdots & x_{t,P,N} \end{bmatrix}, Y_t = \begin{bmatrix} y_{t,1,1} & \cdots & y_{t,Q,1} \\ \vdots & \cdots & \vdots \\ y_{t,1,N} & \cdots & y_{t,Q,N} \end{bmatrix}, M_t = \begin{bmatrix} m_{t,1,1} & \cdots & m_{t,R,1} \\ \vdots & \cdots & \vdots \\ m_{t,1,N} & \cdots & m_{t,R,N} \end{bmatrix}$$

where P is the total number of variables in matrix $X_t$ (i.e. the total number of miRNAs) Q and R are the total number of variables for matrices $Y_t$ and $M_t$ respectively. N denotes the number of participants and the matrices were measured at two time points with $t = 1$ representing pre- and $t = 2$ representing postseason measurements.

Across-season measures for miRNAs, VR scores, and metabolites were calculated as

$$\Delta X = X_2 - X_1$$
$$\Delta Y = Y_2 - Y_1$$
$$\Delta M = M_2 - M_1$$

For this study, permutation-based mediation analysis was performed for all three-way associations following the steps listed below:

1. Mediation analysis was performed by assigning the original data variables $\Delta x_i, \Delta y_j, \Delta m_k$ as independent, dependent, and mediator variables, respectively, to obtain reference the Sobel z-score ($z_0$). Only variables that formed three-way associations were considered. (For pre-season analysis, $x_{1,i}, y_{1,j}, m_{1,k}$ were used to obtain the reference Sobel z-score ($z_0$)).

2. Data permutation: values were randomly selected from $x_{1,i}$ and $x_{2,i}$ to assign to $x'_{1,i}$ and $x'_{2,i}$. (For pre-season analysis, values from $x_{1,i}$ were shuffled to get $x'_{1,i}$.)

3. Across-season measures were computed from the permuted dataset $\Delta x'_i = x'_{2,i} - x'_{1,i}$. (Note that for pre-season analysis, the difference was not computed.)

4. Similarly, $\Delta y'_j$ and $\Delta m'_k$ were computed. (For pre-season analysis, $y'_{1,j}, m'_{1,k}$ were computed by randomly shuffling values in $y_{1,j}, m_{1,k}$.)

5. Mediation analysis was performed on the permuted dataset $\Delta x'_i$, $\Delta y'_j$, $\Delta m'_k$ and the test statistics ($z'_s$) was obtained. (For pre-season analysis, mediation was performed on the permuted data set $x'_{1,i}$, $y'_{1,j}$, $m'_{1,k}$ and $z'_s$ was obtained.)

6. The counter variable $K$ was incremented by one if the absolute value of $z_0$ was greater than the absolute value of $z'_s$.

7. Steps 2-6 were repeated: $s = 1, 2, \cdots, S$ times.

8. Permutation-based *p*-value ($p_{Sobel}^{perm}$) was calculated as the proportion of $z'_s$ values that were as extreme or more extreme than $z_0$ (i.e. $K/S$).

Hypothesis-directed mediation results were considered significant if $p_{Sobel}^{perm} < 0.05$, $T_{eff}$ was $> 50\%$, *and* $T_{eff}$ for the control mediation was $< 30\%$.

Head acceleration event (HAE) monitoring

HAEs were monitored at all contact practice sessions (max = 53) using the BodiTrak sensor system from The Head Health Network [81]. Sensors were mounted in each active player's helmet prior to contact practice (no games were monitored). Sensor outputs included peak translational acceleration (PTA; G-units) and impact location. HAEs were quantified as 1) cumulative hits exceeding 25G and 80G (cHAE$_{25G}$ and cHAE$_{80G}$; Eq. 1) and 2) cumulative hits exceeding 25G and 80G normalized to the total number of sessions per player (aHAE$_{25G}$ and aHAE$_{80G}$; Eq.2). The G-unit thresholds (Th) were selected based on previous reports of impacts related to brain health and injury [82, 83].

$$(1)\ cHAE_{Th,i} = \sum_{k=1}^{N} u(PTA_{k,i} - Th)$$

$$where\ u(x) = \begin{cases} 1\ if\ x > 0 \\ 0\ if\ x \leq 0 \end{cases}$$

$$(2)\ aHAE_{Th,i} = \frac{cHAE_{Th,i}}{sessions_i}$$

RESULTS

**Random forest plot revealed biochemically important metabolites**

Of the 1300+ metabolites analyzed, 20 metabolites with the highest predictive power to distinguish between pre- and postseason (i.e. importance) were plotted (Figure 1). Of these metabolites 11/20 (55%) were lipids, 3/20 (15%) were amino acids, 2/20 (10%) were carbohydrates, 3/20 (15%) were xenobiotics, and 1/20 (5%) was an amino acid. The metabolite with the highest predictive power was 2-hydroxyglutarate (2-HG), followed by three other lipids. All 20 metabolites from the random forest plot were included in additional analyses.

**Wilcoxon signed-rank analyses revealed significant across-season metabolite changes**

Wilcoxon signed-rank tests were conducted to determine which miRNA, metabolites, and VR scores significantly fluctuated between pre- and postseason. There were no significant ($p$-value > 0.05) changes in miRNA levels and VR scores between pre- and postseason. Of the 968 metabolites analyzed, 259 significantly increased or decreased ($p$-value ≤ 0.05) between pre- and postseason. Following FDR correction, 161 significantly increased or decreased ($q$-value ≤ 0.05). Of those 161 metabolites, 40 were selected based on the criteria described in *Metabolomic analysis* (Figure 1). Of the 40 metabolites, 26 significantly increased (negative z-score) and 14 decreased (negative z-score) (Table 1). 48% of the metabolites were lipids and 52% fell in another category (12.5% energy-related, 25% xanthine, 7.5% amino acid, 5% carbohydrate, and 2.5% nucleic acid). Of the lipids, 74% were fatty acids (FAs) and 26% were other lipid types. Overall, 35% of the metabolites were FAs and 65% fell in another category. Of the FAs, 79% significantly increased from pre- to postseason. Additionally, all TCA-related metabolites ($\alpha$-ketoglutarate, citrate, aconitate, and fumarate) decreased from pre- to postseason and all of the xanthine metabolites (e.g. paraxanthine) increased.

**Summary of preseason and across-season interactions between miRNAs, metabolites, and VR scores**

Linear regressions were conducted to assess interactions between four VR scores (Bal, RT, SM, and Comp), nine miRNAs (miR-20a, miR-505, miR-3623p, miR-30d, miR-92a, miR-486, miR-92a, miR-93p, and miR-151-5p), five HAE metrics (sessions, $cHAE_{25G}$, $aHAE_{25G}$, $cHAE_{80G}$, and $aHAE_{80G}$), and 40 metabolites (Supplemental Table 1). At preseason, there were nine significant interactions between VR scores and miRNAs, 15 between VR scores and metabolites, and 36 between miRNAs and metabolites ($p \leq 0.05$ following Cook's outlier removal) (Figure 2). Across-season (postseason−preseason), there were 12 significant interactions between VR scores and miRNAs, 20 between VR scores and metabolites, and 31 between miRNAs and metabolites ($p \leq 0.05$ following Cook's outlier removal) (Figure 3). In general, both pre- and across-season analyses revealed negative relationships between VR scores and miRNAs and positive relationships between VR scores and metabolites (exceptions: cortoloneglucuronide, corticosterone, adenosine, and tridecenedioate). Relationships between miRNAs and metabolites varied with the majority being negative (79%).

# Figure 2: Significant preseason linear regressions

**A**

| VR task (Y) | miRNA (X) | Std. β | $R^2_{adj}$ | p-value | Cook's outliers |
|---|---|---|---|---|---|
| Comprehensive | 20a | -0.631 | 0.251 | 0.014 | 0/20 |
| Comprehensive | 92a | -0.444 | 0.157 | 0.039 | 0/22 |
| Comprehensive | 505 | -0.471 | 0.181 | 0.031 | 0/21 |
| Comprehensive | 30d | -0.605 | 0.333 | 0.004 | 1/22 |
| Comprehensive | 151-5p | -0.661 | 0.390 | 0.001 | 1/22 |
| Balance | 505 | -0.721 | 0.476 | 0.000 | 1/21 |
| Balance | 151-5p | -0.579 | 0.294 | 0.012 | 4/22 |
| Reaction Time | 195 | -0.472 | 0.182 | 0.031 | 1/22 |
| Spatial Memory | 486 | -0.546 | 0.215 | 0.013 | 2/22 |

**B**

| VR task (Y) | Metabolite (X) | Std. β | $R^2_{adj}$ | p-value | Cook's outliers |
|---|---|---|---|---|---|
| Comprehensive | 2-hydroxyglutarate | 0.760 | 0.558 | 0.000 | 0/23 |
| Comprehensive | cortisol | 0.484 | 0.191 | 0.031 | 3/23 |
| Comprehensive | sebacate | 0.523 | 0.237 | 0.013 | 1/23 |
| Comprehensive | 8-hydroxyoctanate | 0.612 | 0.343 | 0.002 | 1/23 |
| Comprehensive | undecanedioate | 0.528 | 0.242 | 0.012 | 1/23 |
| Balance | 2-hydroxyglutarate | 0.475 | 0.183 | 0.034 | 3/23 |
| Balance | cortoloneglucuronide | -0.536 | 0.249 | 0.012 | 2/23 |
| Balance | sebacate | 0.498 | 0.208 | 0.022 | 2/23 |
| Balance | suberate | 0.425 | 0.140 | 0.049 | 1/23 |
| Balance | 8-hydroxyoctanate | 0.560 | 0.278 | 0.008 | 2/23 |
| Balance | undecanedioate | 0.652 | 0.393 | 0.002 | 3/23 |
| Reaction Time | 2-hydroxyglutarate | 0.525 | 0.241 | 0.010 | 0/23 |
| Reaction Time | sebacate | 0.516 | 0.229 | 0.014 | 1/23 |
| Reaction Time | 8-hydroxyoctanate | 0.503 | 0.218 | 0.014 | 0/23 |
| Reaction Time | undecanedioate | 0.547 | 0.262 | 0.010 | 2/23 |

**C**

| Metabolite (Y) | miRNA (X) | Std. β | $R^2_{adj}$ | p-value | Cook's outliers |
|---|---|---|---|---|---|
| 2-hydroxglutarate | 20a | -0.529 | 0.235 | 0.024 | 2/20 |
| sebacate | 20a | -0.486 | 0.191 | 0.035 | 1/20 |
| 8-hydroxyoctanate | 20a | -0.619 | 0.347 | 0.005 | 1/20 |
| undecanedioate | 20a | -0.589 | 0.306 | 0.010 | 2/20 |
| heptanoate (7:0) | 20a | -0.470 | 0.175 | 0.042 | 1/20 |
| citrate | 505 | -0.450 | 0.156 | 0.050 | 2/21 |
| 2-hydroxyglutarate | 505 | -0.657 | 0.400 | 0.002 | 1/21 |
| 8-hydroxyoctanate | 505 | -0.508 | 0.219 | 0.019 | 1/21 |
| undecanedioate | 505 | -0.478 | 0.183 | 0.038 | 2/21 |
| 1-carboxyethylphenylalanine | 505 | -0.514 | 0.218 | 0.029 | 3/21 |
| heptanoate (7:0) | 505 | -0.458 | 0.168 | 0.037 | 0/21 |
| cortisone | 3623p | -0.567 | 0.286 | 0.007 | 1/22 |
| phosphate | 3623p | 0.572 | 0.287 | 0.011 | 3/22 |
| paraxanthine | 3623p | 0.716 | 0.485 | 0.000 | 2/22 |
| 1,7-dimethylurate | 3623p | 0.459 | 0.167 | 0.042 | 2/22 |
| 1-methylxanthine | 3623p | 0.513 | 0.222 | 0.021 | 2/22 |
| 1 2 3-benzenetriol sulfate | 3623p | 0.721 | 0.495 | 0.000 | 1/22 |
| cortisone | 30d | -0.509 | 0.223 | 0.015 | 0/22 |
| cortoloneglucuronide | 30d | 0.470 | 0.229 | 0.032 | 1/22 |
| 2-hydroxyglutarate | 30d | -0.450 | 0.156 | 0.053 | 3/22 |
| heptanoate (7:0) | 30d | -0.445 | 0.158 | 0.038 | 0/22 |
| citrate | 92a | -0.473 | 0.185 | 0.026 | 0/22 |
| 8-hydroxyoctanate | 92a | -0.552 | 0.268 | 0.010 | 1/22 |
| heptanoate (7:0) | 92a | -0.567 | 0.284 | 0.009 | 2/22 |
| cortoloneglucuronide | 486 | -0.544 | 0.257 | 0.013 | 2/22 |
| citrate | 195 | -0.486 | 0.198 | 0.022 | 0/22 |
| afm | 195 | 0.248 | 0.210 | 0.028 | 3/22 |
| 8-hydroxyoctanate | 195 | -0.443 | 0.156 | 0.039 | 0/22 |
| heptanoate (7:0) | 195 | -0.538 | 0.252 | 0.012 | 1/22 |
| 1 2 3-benzenetriol sulfate | 195 | 0.740 | 0.522 | 0.000 | 2/22 |
| aconitate | 151-5p | -0.429 | 0.137 | 0.051 | 0/22 |
| 2-hydroxyglutarate | 151-5p | -0.516 | 0.226 | 0.02 | 2/22 |
| azelate | 151-5p | -0.445 | 0.156 | 0.043 | 1/22 |
| 8-hydroxyoctanate | 151-5p | -0.554 | 0.270 | 0.009 | 1/22 |
| undecanedioate | 151-5p | -0.575 | 0.293 | 0.008 | 2/22 |
| 1-carboxyethylphenylalanine | 151-5p | -0.490 | 0.198 | 0.028 | 2/22 |

**Figure 2:** Preseason linear regression results. Cook's outliers were removed prior to regression analysis and the number of outliers removed in each regression is listed as the ratio of outliers to the total n. Adjusted $R^2$ ($R^2_{adj.}$) is reported for each regression and relays the amount of variance in Y that can be explained by X. p-Values are reported at a significance level of 0.05. **(A)** Significant interactions between VR scores and miRNA levels. **(B)** Significant interactions between VR scores and metabolite levels. **(C)** Significant interactions between miRNA and metabolite levels.

## Figure 3: Significant across season linear regressions

**A**

| VR task (Y) | miRNA (X) | Std. β | $R^2_{adj}$ | p-value | Cook's outliers |
|---|---|---|---|---|---|
| Comprehensive | 505 | -0.649 | 0.387 | 0.003 | 1/20 |
| Comprehensive | 30d | -0.586 | 0.306 | 0.007 | 0/20 |
| Comprehensive | 92a | -0.479 | 0.186 | 0.033 | 0/20 |
| Comprehensive | 195 | -0.436 | 0.145 | 0.050 | 0/20 |
| Comprehensive | 151-5p | -0.655 | 0.395 | 0.002 | 1/20 |
| Balance | 505 | -0.651 | 0.389 | 0.003 | 1/20 |
| Balance | 30d | -0.507 | 0.213 | 0.027 | 1/20 |
| Reaction Time | 20a | -0.482 | 0.185 | 0.043 | 0/18 |
| Reaction Time | 505 | -0.544 | 0.254 | 0.016 | 1/20 |
| Reaction Time | 30d | -0.483 | 0.190 | 0.031 | 0/20 |
| Reaction Time | 92a | -0.537 | 0.249 | 0.015 | 0/20 |
| Reaction Time | 151-5p | -0.651 | 0.323 | 0.007 | 1/20 |

**B**

| VR task (Y) | Metabolite (X) | Std. β | $R^2_{adj}$ | p-value | Cook's outliers |
|---|---|---|---|---|---|
| Comprehensive | 2-hydroxyglutarate | 0.561 | 0.279 | 0.008 | 2/23 |
| Comprehensive | corticosterone | -0.563 | 0.281 | 0.008 | 2/23 |
| Comprehensive | 1,7-dimethylurate | 0.452 | 0.165 | 0.035 | 1/23 |
| Comprehensive | sebacate | 0.636 | 0.373 | 0.002 | 2/23 |
| Comprehensive | dodecadienoate | 0.452 | 0.160 | 0.045 | 3/23 |
| Comprehensive | azelate | 0.484 | 0.196 | 0.022 | 1/23 |
| Comprehensive | suberate | 0.502 | 0.214 | 0.017 | 1/23 |
| Comprehensive | 8-hydroxyoctanate | 0.558 | 0.276 | 0.007 | 1/23 |
| Comprehensive | heptanoate (7:0) | 0.532 | 0.248 | 0.011 | 1/23 |
| Comprehensive | 1 2 3-benzenetriol sulfate | 0.467 | 0.179 | 0.029 | 1/23 |
| Comprehensive | adenosine | -0.609 | 0.341 | 0.002 | 0/23 |
| Balance | azelate | 0.461 | 0.174 | 0.031 | 1/23 |
| Balance | 7-hydroxyoctanate | 0.451 | 0.162 | 0.040 | 2/23 |
| Balance | 8-hydroxyoctanate | 0.607 | 0.335 | 0.004 | 2/23 |
| Balance | undecanedioate | 0.426 | 0.140 | 0.048 | 1/23 |
| Balance | heptanoate (7:0) | 0.455 | 0.167 | 0.033 | 1/23 |
| Balance | tridecenedioate | -0.439 | 0.152 | 0.041 | 1/23 |
| Reaction Time | alpha-ketoglutarate | 0.483 | 0.193 | 0.027 | 2/23 |
| Reaction Time | 2-hydroxyglutarate | 0.541 | 0.256 | 0.011 | 2/23 |
| Reaction Time | adenosine | -0.591 | 0.317 | 0.004 | 2/23 |

**C**

| Metabolite (Y) | miRNA (X) | Std. β | $R^2_{adj}$ | p-value | Cook's outliers |
|---|---|---|---|---|---|
| azelate | 20a | -0.636 | 0.365 | 0.006 | 1/18 |
| sebacate | 505 | -0.591 | 0.309 | 0.010 | 2/20 |
| azelate | 505 | -0.662 | 0.404 | 0.003 | 20/20 |
| suberate | 505 | -0.600 | 0.320 | 0.008 | 2/20 |
| 8-hydroxyoctanate | 505 | -0.485 | 0.188 | 0.041 | 2/20 |
| undecanedioate | 505 | -0.604 | 0.326 | 0.008 | 2/20 |
| heptanoate (7:0) | 505 | -0.514 | 0.218 | 0.029 | 2/20 |
| tridecenedioate | 505 | 0.450 | 0.156 | 0.050 | 1/20 |
| phosphate | 3623p | -0.495 | 0.163 | 0.050 | 2/20 |
| linoleate3n6 | 3623p | 0.509 | 0.210 | 0.037 | 3/20 |
| heptanoate (7:0) | 3623p | -0.543 | 0.256 | 0.013 | 0/20 |
| sebacate | 30d | -0.528 | 0.239 | 0.017 | 0/20 |
| heptanoate (7:0) | 30d | -0.618 | 0.346 | 0.005 | 1/20 |
| adenosine | 30d | 0.550 | 0.259 | 0.018 | 2/20 |
| sebacate | 92a | -0.734 | 0.510 | 0.001 | 2/20 |
| azelate | 92a | -0.689 | 0.437 | 0.003 | 4/20 |
| suberate | 92a | -0.622 | 0.346 | 0.008 | 3/20 |
| undecanedioate | 92a | -0.641 | 0.375 | 0.004 | 2/20 |
| heptanoate (7:0) | 92a | -0.523 | 0.234 | 0.018 | 0/20 |
| O-sulfo-L-tyrosine | 92a | -0.506 | 0.215 | 0.023 | 0/20 |
| adenosine | 92a | 0.496 | 0.199 | 0.036 | 2/20 |
| sebacate | 195 | -0.545 | 0.256 | 0.016 | 1/20 |
| suberate | 195 | -0.476 | 0.175 | 0.053 | 3/20 |
| 8-hydroxyoctanate | 195 | -0.632 | 0.364 | 0.004 | 1/20 |
| undecanedioate | 195 | -0.530 | 0.239 | 0.020 | 1/20 |
| heptanoate (7:0) | 195 | -0.667 | 0.412 | 0.002 | 1/20 |
| O-sulfo-L-tyrosine | 195 | -0.530 | 0.238 | 0.020 | 1/20 |
| 7-hydroxyoctanate | 93p | 0.535 | 0.247 | 0.018 | 1/20 |
| stearidonate | 151-5p | 0.556 | 0.263 | 0.020 | 3/20 |
| 8-hydroxyoctanate | 151-5p | -0.480 | 0.185 | 0.038 | 1/20 |
| heptanoate (7:0) | 151-5p | -0.562 | 0.273 | 0.015 | 2/20 |

**Figure 3**: Across-season linear regression results. Variables are expressed as the difference between postseason and preseason values (post-pre = Δ). Cook's outliers were removed prior to regression analysis and the number of outliers removed in each regression is listed as the ratio of outliers to the total n. Adjusted $R^2$ ($R^2_{adj.}$) is reported for each regression and relays the amount of variance in Y that can be explained by X. *p*-Values are reported at a significance level of 0.05. **(A)** Significant interactions between ΔVR scores and ΔmiRNA levels (IV). **(B)** Significant interactions between ΔVR scores and Δmetabolite levels. **(C)** Significant interactions between ΔmiRNA and Δmetabolite levels.

**Three-way associations between miRNAs, metabolites, and VR scores at preseason and across-season**

As a prerequisite for mediation analysis, only regressions which formed three-way associations were considered (e.g. significant interactions between X → Y, Y → Z, and X → Z; see below).

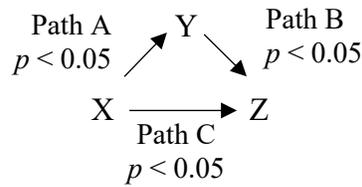

In total, there were 50 potential three-way associations – 18 at preseason and 32 across-season. The majority of three-way associations included Comp behavior (70%) and 88% of three-way associations included a FA metabolomic measure. Following Cook's outlier removal across all regressions, 11/18 preseason three-way associations met significance ($p$-value $< 0.05$) for all pathways (A, B, and C). Across-season, 14/32 associations met a $p$-value $< 0.05$ threshold. Of the 25 significant three-way associations, 92% included a FA metabolomic measure and 88% included Comp behavior. There were six miRNAs involved in pre- and across-season three-way associations: miR-20a, miR-505, miR-92a, miR-151-5p, miR-195, and miR-30d (five at preseason and five across-season).

**Mediation analyses revealed causal mediations at preseason and across-season**

Mediation analyses were performed to assess the causal influence of metabolites on the relationship between miRNAs and VR scores and contrasted against a control mediation analyses that switched the roles of metabolites and mRNAs. Specifically, the hypothesized mediation was directed so miRNA was the independent variable, metabolite was the mediator, and VR score was the dependent variable. Of the 18 preseason three-way associations, 11 survived following combined removal of Cook's outliers across all paths (A, B, and C) – ensuring the same set of subjects were being analyzed. Mediation analysis revealed six significant preseason mediations (Sobel $p$-value $< 0.05$, $T_{eff} \geq 50\%$; Table 2, bolded and italicized rows). Figure 4 depicts each significant mediation, including the directionality and significance of each regression interaction.

The graphs illustrate the slopes of linear models 1 (IV → DV) versus 2 (IV + M → DV). In each case, the slope of model 1 was larger than that of model 2, indicating that the metabolite (M) significantly impacted the relationship between the miRNA (IV) and VR score (DV). The directionality of the linear regression relationships was common across all preseason mediations – negative between miRNA and metabolite, negative between miRNA and VR score, and positive between metabolite and VR score. No control mediations met both a significant *p*-value and a T$_{eff}$ greater than 50% when metabolite was the IV, miRNA was the M, and VR score was the DV (*p*-value > 0.048, T$_{eff}$ ≤ 30%).

**Table 2: Preseason mediation analysis**

| | Model 1: IV Path A -> M Path B -> DV<br>Model 2: IV Path C -> DV | | | Path A: IV predicting M | | Path B: M predicting DV | | Path C: IV predicting DV (model 1) | | Path C: IV (with M) predicting DV (model 2) | | Effect Mediated | $p^{perm}_{Sobel}$ |
|---|---|---|---|---|---|---|---|---|---|---|---|---|---|
| VR Score (DV) | Mediation | IV | M | Std β | $p_a$ | Std β | $p_b$ | Std β | $p_c$ | Std β | p | % | |
| Comprehensive | *Hypothesis* | *miR-20a* | *2-hydroxyglutarate* | *-0.529* | *0.024* | *0.655* | *0.003* | *-0.531* | *0.023* | *-0.256* | *0.263* | *52* | *0.002* |
| | Control | 2-hydroxyglutarate | miR-20a | -0.529 | 0.024 | -0.531 | 0.023 | 0.655 | 0.003 | 0.520 | 0.032 | 21 | 0.048 |
| Comprehensive | Hypothesis | miR-20a | sebacate | -0.489 | 0.039 | 0.538 | 0.021 | -0.567 | 0.014 | -0.400 | 0.099 | 30 | 0.054 |
| | Control | sebacate | miR-20a | -0.514 | 0.035 | -0.447 | 0.072 | 0.498 | 0.042 | 0.364 | 0.184 | 27 | 0.103 |
| Comprehensive | *Hypothesis* | *miR-20a* | *8-hydroxyoctanoate* | *-0.631* | *0.005* | *0.648* | *0.004* | *-0.544* | *0.020* | *-0.224* | *0.378* | *59* | *0.005* |
| | Control | 8-hydroxyoctanoate | miR-20a | -0.508 | 0.037 | -0.413 | 0.100 | 0.577 | 0.015 | 0.495 | 0.068 | 14 | 0.269 |
| Comprehensive | Hypothesis | miR-20a | undecanedionate | -0.630 | 0.005 | 0.622 | 0.006 | -0.575 | 0.013 | -0.304 | 0.239 | 32 | 0.113 |
| | Control | undecanedionate | miR-20a | -0.644 | 0.005 | -0.390 | 0.122 | 0.438 | 0.079 | 0.320 | 0.32 | 27 | 0.297 |
| Comprehensive | *Hypothesis* | *miR-505* | *2-hydroxyglutarate* | *-0.657* | *0.002* | *0.737* | *0.000* | *-0.513* | *0.021* | *-0.050* | *0.819* | *90* | *0.000* |
| | Control | 2-hydroxyglutarate | miR-505 | -0.491 | 0.033 | -0.363 | 0.126 | 0.680 | 0.001 | 0.661 | 0.006 | 3 | 0.747 |
| Comprehensive | Hypothesis | miR-505 | 8-hydroxyoctanoate | -0.475 | 0.034 | 0.639 | 0.002 | -0.515 | 0.020 | -0.273 | 0.193 | 47 | 0.007 |
| | Control | 8-hydroxyoctanoate | miR-505 | -0.292 | 0.225 | -0.360 | 0.130 | 0.573 | 0.010 | 0.511 | 0.026 | 11 | 0.148 |
| Comprehensive | *Hypothesis* | *miR-92a* | *8-hydroxyoctanoate* | *-0.551* | *0.012* | *0.688* | *0.001* | *-0.447* | *0.048* | *-0.098* | *0.648* | *78* | *0.001* |
| | Control | 8-hydroxyoctanoate | miR-92a | -0.271 | 0.277 | -0.236 | 0.345 | 0.613 | 0.007 | 0.592 | 0.013 | 3 | 0.516 |
| Comprehensive | Hypothesis | miR-151-5p | 2-hydroxyglutarate | -0.533 | 0.019 | 0.718 | 0.001 | -0.612 | 0.005 | -0.320 | 0.11 | 48 | 0.000 |
| | Control | 2-hydroxyglutarate | miR-151-5p | -0.572 | 0.010 | -0.582 | 0.009 | 0.747 | 0.000 | 0.616 | 0.006 | 18 | 0.061 |
| Comprehensive | *Hypothesis* | *miR-151-5p* | *8-hydroxyoctanoate* | *-0.572* | *0.010* | *0.689* | *0.001* | *-0.595* | *0.007* | *-0.298* | *0.171* | *50* | *0.004* |
| | Control | 8-hydroxyoctanoate | miR-151-5p | -0.588 | 0.008 | -0.665 | 0.002 | 0.724 | 0.001 | 0.509 | 0.018 | 30 | 0.009 |
| Balance | *Hypothesis* | *miR-151-5p* | *undecanedionate* | *-0.575* | *0.016* | *0.757* | *0.000* | *-0.578* | *0.015* | *-0.213* | *0.317* | *63* | *0.001* |
| | Control | undecanedionate | miR-151-5p | -0.575 | 0.016 | -0.578 | 0.015 | 0.757 | 0.000 | 0.635 | 0.008 | 16 | 0.085 |
| Reaction Time | Hypothesis | miR-195 | 8-hydroxyoctanoate | -0.458 | 0.037 | 0.553 | 0.009 | -0.472 | 0.031 | -0.277 | 0.205 | 41 | 0.014 |
| | Control | 8-hydroxyoctanoate | miR-195 | -0.118 | 0.631 | -0.338 | 0.157 | 0.471 | 0.042 | 0.437 | 0.054 | 7 | 0.391 |

**Table 2:** Permutation-based preseason mediation results. The *a priori* hypothesis assumed that miRNA was the IV, metabolite was the M, and VR score was the DV. Control mediations were run with metabolite as the IV, miRNA as the M, and VR score as the DV. Beta coefficients and standard error terms from Table 1 were used to calculate permutation-based Sobel *p*-values (significance level = 0.05) and total effect mediated (T$_{eff}$; expressed as %). Results with permutation-based Sobel *p*-values ($p^{perm}_{Sobel}$) ≥ 0.05 and % ≥ 50 were considered significant. Of the 11 three-way associations, six showed a significant mediation effect (***bolded and italicized rows***).

**Figure 4: Preseason mediations**

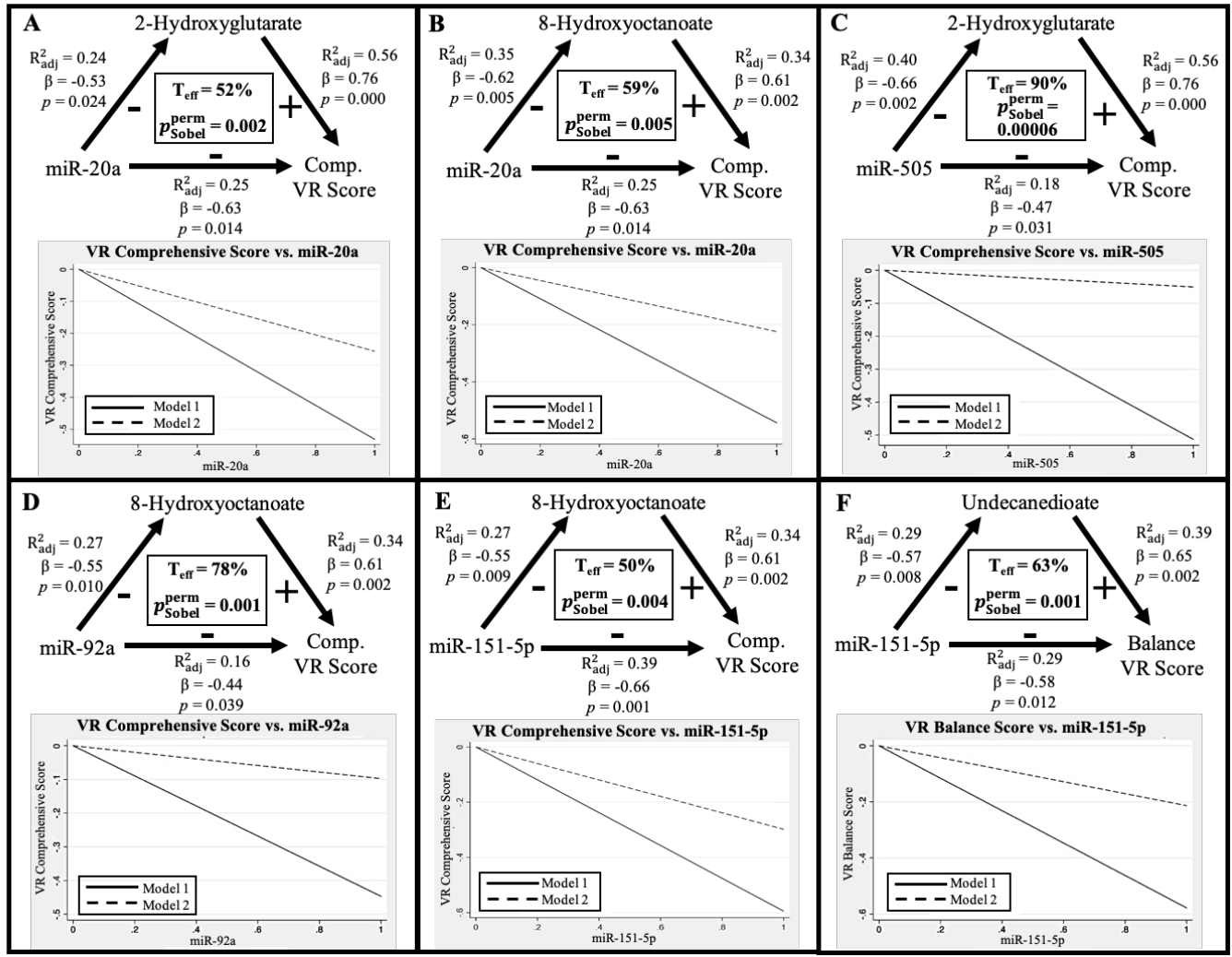

**Figure 4:** Significant preseason mediation results. In all analyses, miRNA was the independent variable, metabolite was the mediator, and VR score was the dependent variable. Cook's distance outliers were removed prior to analyses, VR terms were reported as standardized values, Adjusted $R^2$ ($R^2_{adj.}$) values were reported for each regression, and all *p*-values were reported at a significance level 0.05. **(A)** There was a negative interaction between miR-20a and 2-hydroxyglutarate (2-HG), a positive interaction between 2-HG and VR composite score (Comp), and a negative interaction between miR-20a and Comp. When 2-HG was added to the regression model, the relationship between miR-20a and Comp no longer existed (*p*-value > 0.05); therefore, 2-HG significantly mediated the relationship (Sobel *p*-value = 0.002, $T_{eff}$ = 52%). The graph depicts the change in slope between model 1, which plots the slope term for the interaction between miR-20a and Comp, and model 2, which plots the slope term for the interaction between miR-20a and Comp when 2-HG was included in the regression model. **(B)** There was a negative interaction between miR-20a and 8-hydroxyoctanoate (8-HOA), a positive interaction between 8-HOA and Comp, and a negative interaction between miR-20a and Comp. When 8-HOA was added to the regression model, the relationship between miR-20a and Comp no longer existed (*p*-value > 0.05); therefore, 8-HOA significantly mediated the relationship (Sobel *p*-value = 0.005, $T_{eff}$ = 59%). The graph depicts the change in slope between model 1, which plots the slope term for the interaction between miR-20a and Comp, and model 2, which plots the slope term for the interaction between miR-20a and Comp when 8-HOA was included in the regression model. **(C)** There was a significant negative interaction between miR-505 and 2-HG, a significant positive interaction between 2-HG and Comp, and a significant negative interaction between miR-505 and Comp. When 2-HG was added to the regression model, the relationship between miR-505 and Comp no longer existed ($p$-value > 0.05); therefore, 2-HG significantly mediated the relationship (Sobel $p$-value = 0.00006, $T_{eff}$ = 90%). The graph depicts the change in slope between model 1, which plots the slope term for the interaction between miR-505 and Comp, and model 2, which plots the slope term for the interaction between miR-505 and Comp when 2-HG was included in the regression model. **(D)** There was a negative interaction between miR-92a and 8-HOA, a positive interaction between 8-HOA and Comp, and a negative interaction between miR-92a and Comp. When 8-HOA was added to the regression model, the relationship between miR-92a and Comp no longer existed ($p$-value > 0.05); therefore, 8-HOA significantly mediated the relationship (Sobel $p$-value = 0.001, $T_{eff}$ = 78%). The graph depicts the change in slope between model 1, which plots the slope term for the interaction between miR-92a and Comp, and model 2, which plots the slope term for the interaction between miR-92a and Comp when 8-HOA was included in the regression model. **(E)** There was a negative interaction between miR-151-5p and 8-HOA, a positive interaction between 8-HOA and Comp, and a negative interaction between miR-151-5p and Comp. When 8-HOA was added to the regression model, the relationship between miR-151-5p and Comp no longer existed ($p$-value > 0.05); therefore, 8-HOA significantly mediated the relationship (Sobel $p$-value = 0.004, $T_{eff}$ = 50%). The graph depicts the change in slope between model 1, which plots the slope term for the interaction between miR-151-5p and Comp, and model 2, which plots the slope term for the interaction between miR-151-5p and Comp when 8-HOA was included in the regression model. **(F)** There was a negative interaction between miR-151-5p and undecanedioate (UND), a positive interaction between UND and Bal, and a negative interaction between miR-151-5p and Bal. When UND was added to the regression model, the relationship between miR-151-5p and Bal no longer existed ($p$-value > 0.05); therefore, UND significantly mediated the relationship (Sobel $p$-value = 0.001, $T_{eff}$ = 63%). The graph depicts the change in slope between model 1, which plots the slope term for the interaction between miR-151-5p and Bal, and model 2, which plots the slope term for the interaction between miR-151-5p and Bal when UND was included in the regression model.

Across-season mediation analysis revealed eight mediations meeting our thresholds (Sobel $p$-value < 0.05; $T_{eff}$ > 50%; Table 3, bolded and italicized rows). Figure 5A-H visualizes each across-season mediation, including the directionality of each interaction. The plots depict the slopes of linear models 1 (IV → DV) versus 2 (IV + M → DV). In each case, the slope of model 1 was larger than that of model 2. Mediations depicted in Figures 5A-C and 5F-H share the same directionality across regressions: negative between miRNA and metabolite, positive between metabolite and VR score, and negative between miRNA and VR score. In 5D-E, the interactions between miRNA (miR-30d) and metabolite (adenosine), as well as metabolite and VR score (Comp), were opposite of those depicted in 5A-C. Only one control mediation met threshold when metabolite was the IV, miRNA was the M, and VR score was the DV ($p$-value = 0.011, $T_{eff}$ =

68%); in this case the hypothesized mediation also had a significant Sobel p-value (*p*-value = 0.047), indicating a mixed mediation model.

**Table 3: Across-season mediation analysis**

| VR Score (DV) | Mediation | IV | M | Path A: IV predicting M Std β | $p_a$ | Path B: M predicting DV Std β | $p_b$ | Path C: IV predicting DV (model 1) Std β | $p_c$ | Path C: IV (with M) predicting DV (model 2) Std β | $p$ | Effect Mediated % | $p_{Sobel}^{perm}$ |
|---|---|---|---|---|---|---|---|---|---|---|---|---|---|
| Comprehensive | *Hypothesis* | *miR-505* | *sebacate* | *-0.605* | *0.010* | *0.655* | *0.004* | *-0.579* | *0.015* | *-0.289* | *0.252* | *50* | *0.008* |
|  | Control | sebacate | miR-505 | -0.569 | 0.017 | -0.531 | 0.028 | 0.629 | 0.007 | 0.483 | 0.067 | 23 | 0.089 |
| Comprehensive | Hypothesis | miR-505 | azelate | -0.643 | 0.005 | 0.626 | 0.007 | -0.579 | 0.015 | -0.302 | 0.265 | 48 | 0.017 |
|  | Control | azelate | miR-505 | -0.670 | 0.004 | -0.524 | 0.037 | 0.630 | 0.009 | 0.506 | 0.100 | 20 | 0.256 |
| Comprehensive | Hypothesis | miR-505 | suberate | -0.638 | 0.006 | 0.592 | 0.012 | -0.579 | 0.015 | -0.341 | 0.219 | 41 | 0.035 |
|  | Control | suberate | miR-505 | -0.621 | 0.008 | -0.531 | 0.028 | 0.568 | 0.017 | 0.388 | 0.172 | 32 | 0.073 |
| Comprehensive | *Hypothesis* | *miR-505* | *8-hydroxyoctanoate* | *-0.526* | *0.037* | *0.698* | *0.003* | *-0.524* | *0.037* | *-0.217* | *0.353* | *59* | *0.007* |
|  | Control | 8-hydroxyoctanoate | miR-505 | -0.526 | 0.037 | -0.524 | 0.037 | 0.698 | 0.003 | 0.584 | 0.023 | 16 | 0.123 |
| Comprehensive | *Hypothesis* | *miR-30d* | *sebacate* | *-0.595* | *0.007* | *0.662* | *0.002* | *-0.586* | *0.008* | *-0.298* | *0.196* | *49* | *0.003* |
|  | Control | sebacate | miR-30d | -0.457 | 0.056 | -0.472 | 0.048 | 0.594 | 0.009 | 0.478 | 0.050 | 19 | 0.074 |
| Comprehensive | Hypothesis | miR-30d | suberate | -0.518 | 0.033 | 0.627 | 0.007 | -0.551 | 0.022 | -0.309 | 0.199 | 44 | 0.009 |
|  | Control | suberate | miR-30d | -0.515 | 0.029 | -0.472 | 0.048 | 0.552 | 0.018 | 0.420 | 0.103 | 24 | 0.066 |
| Comprehensive | Hypothesis | miR-30d | heptanoate | -0.618 | 0.005 | 0.619 | 0.005 | -0.586 | 0.008 | -0.328 | 0.183 | 44 | 0.009 |
|  | Control | heptanoate | miR-30d | -0.618 | 0.005 | -0.586 | 0.008 | 0.619 | 0.005 | 0.416 | 0.097 | 48 | 0.034 |
| Comprehensive | *Hypothesis* | *miR-30d* | *adenosine* | *0.550* | *0.018* | *-0.620* | *0.006* | *-0.474* | *0.047* | *-0.191* | *0.434* | *60* | *0.001* |
|  | Control | adenosine | miR-30d | 0.550 | 0.018 | -0.474 | 0.047 | -0.620 | 0.006 | -0.515 | 0.047 | 17 | 0.124 |
| Comprehensive | *Hypothesis* | *miR-92a* | *suberate* | *-0.622* | *0.008* | *0.653* | *0.004* | *-0.505* | *0.039* | *-0.161* | *0.537* | *68* | *0.004* |
|  | Control | suberate | miR-92a | -0.330 | 0.196 | -0.286 | 0.266 | 0.578 | 0.015 | 0.542 | 0.033 | 6 | 0.403 |
| Comprehensive | *Hypothesis* | *miR-92a* | *heptanoate* | *-0.523* | *0.018* | *0.581* | *0.007* | *-0.479* | *0.033* | *-0.240* | *0.299* | *50* | *0.006* |
|  | Control | heptanoate | miR-92a | -0.383 | 0.129 | -0.376 | 0.137 | 0.406 | 0.106 | 0.306 | 0.250 | 24 | 0.115 |
| Comprehensive | *Hypothesis* | *miR-195* | *heptanoate* | *-0.667* | *0.002* | *0.530* | *0.019* | *-0.491* | *0.033* | *-0.247* | *0.386* | *50* | *0.029* |
|  | Control | heptanoate | miR-195 | -0.542 | 0.020 | -0.292 | 0.240 | 0.415 | 0.087 | 0.363 | 0.212 | 12 | 0.482 |
| Comprehensive | *Hypothesis* | *miR-151-5p* | *8-hydroxyoctanoate* | *-0.653* | *0.006* | *0.698* | *0.003* | *-0.616* | *0.011* | *-0.279* | *0.286* | *55* | *0.007* |
|  | Control | 8-hydroxyoctanoate | miR-151-5p | -0.532 | 0.034 | -0.536 | 0.032 | 0.676 | 0.004 | 0.545 | 0.035 | 19 | 0.087 |
| Comprehensive | Hypothesis | miR-151-5p | heptanoate | -0.585 | 0.014 | 0.555 | 0.021 | -0.569 | 0.017 | -0.371 | 0.169 | 35 | 0.047 |
|  | Control | heptanoate | miR-151-5p | -0.678 | 0.004 | -0.604 | 0.013 | 0.495 | 0.051 | 0.158 | 0.604 | 68 | 0.011 |
| Reaction Time | *Hypothesis* | *miR-30d* | *adenosine* | *0.529* | *0.029* | *-0.622* | *0.008* | *-0.473* | *0.055* | *-0.200* | *0.420* | *58* | *0.001* |
|  | Control | adenosine | miR-30d | 0.529 | 0.029 | -0.473 | 0.055 | -0.622 | 0.008 | -0.516 | 0.050 | 17 | 0.124 |

**Table 3:** Permutation-based across-season mediation results. The *a priori* hypothesis assumed that ΔmiRNA was the IV, Δmetabolite was the M, and ΔVR score was the DV. Control mediations were run with Δmetabolite as the IV, ΔmiRNA as the M, and ΔVR score as the DV. Beta coefficients and standard error terms from Table 1 were used to calculate permutation-based Sobel *p*-values (significance level = 0.05) and total effect mediated (T$_{eff}$; expressed as %). Results with permutation-based Sobel *p*-values ($p_{Sobel}^{perm}$) ≥ 0.05 and % ≥ 50 were considered significant. Of the 14 three-way associations, eight showed a significant mediation effect (**bolded and italicized rows**).

**Figure 5A-C: Across-season mediations with significant HAE regressions (Post-Pre)**

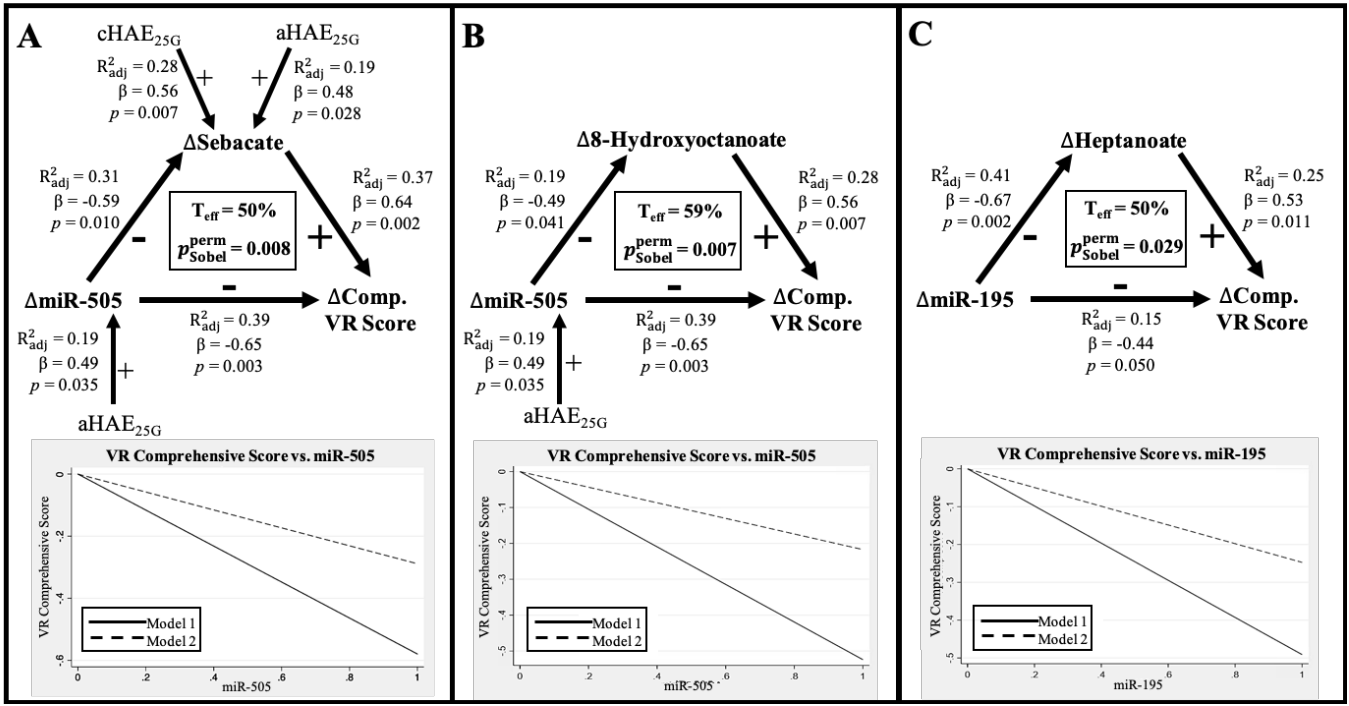

**Figure 5D-F: Across-season mediations with significant HAE regressions (Post-Pre)**

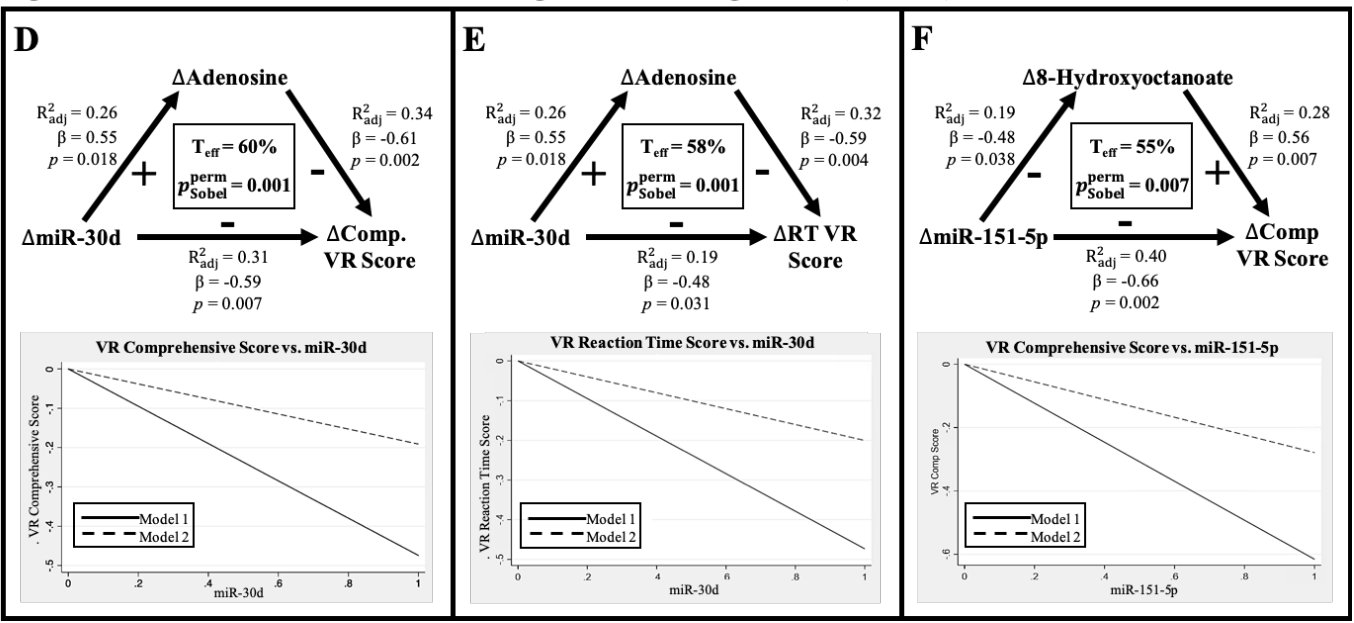

**Figure 5G-H: Across-season mediations (Post-Pre)**

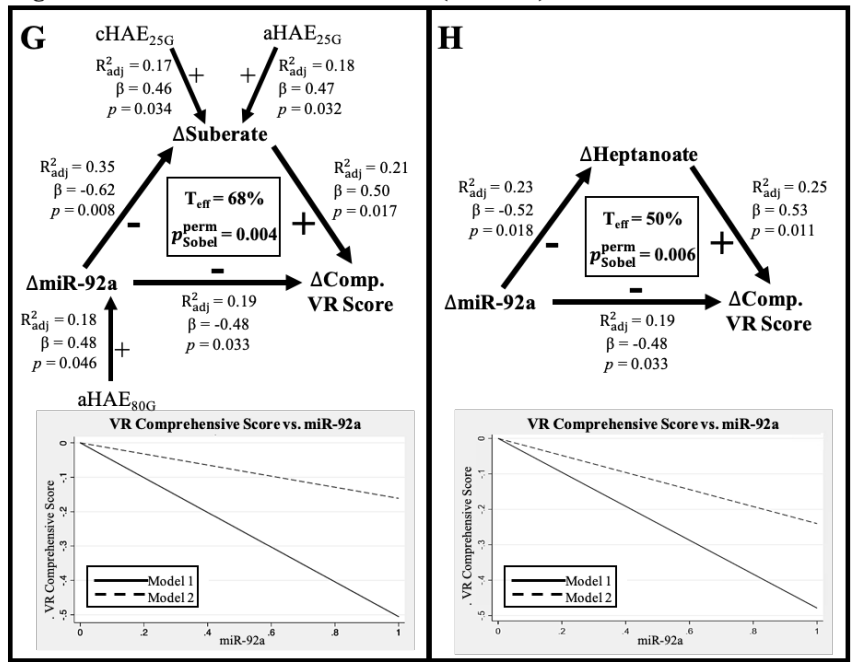

**Figure 5A-H**: Significant across-season mediation results. In all analyses, ΔmiRNA was the independent variable, Δmetabolite was the mediator, and ΔVR score was the dependent variable. Cook's distance outliers were removed prior to analyses, ΔVR terms were reported as standardized values, Adjusted $R^2$ ($R^2_{adj.}$) values were reported for each regression, and all *p*-values were reported at a significance level 0.05. **(A)** There was a negative interaction between ΔmiR-505 and Δsebacate, a positive interaction between Δsebacate and ΔComp, and a negative interaction between ΔmiR-505 and ΔComp. When Δsebacate was added to the regression model, the relationship between ΔmiR-505 and ΔComp no longer existed; therefore, sebacate significantly mediated the relationship (Sobel *p*-value = 0.008, $T_{eff}$ = 50%). The graph depicts the change in slope between model 1, which plots the slope term for the interaction between ΔmiR-505 and ΔComp, and model 2, which plots the slope term for the interaction between ΔmiR-505 and ΔComp when Δsebacate was included in the regression model. **(B)** There was a negative interaction between ΔmiR-505 and Δ8-hydroxyoctanoate (Δ8-HOA), a positive interaction between Δ8-HOA and ΔComp, and a negative interaction between ΔmiR-505 and ΔComp. When Δ8-HOA was added to the regression model, the relationship between ΔmiR-505 and ΔComp no longer existed; therefore, 8-HOA significantly mediated the relationship (Sobel *p*-value = 0.007, $T_{eff}$ = 59%). The graph depicts the change in slope between model 1, which plots the slope term for the interaction between ΔmiR-505 and ΔComp, and model 2, which plots the slope term for the interaction between ΔmiR-505 and ΔComp when Δ8-HOA was included in the regression model. **(C)** There was a negative interaction between ΔmiR-195 and Δheptanoate, a positive interaction between Δheptanoate and ΔComp, and a negative interaction between ΔmiR-195 and ΔComp. When Δheptanoate was added to the regression model, the relationship between ΔmiR-195 and ΔComp no longer existed; therefore, heptanoate significantly mediated the relationship (Sobel *p*-value = 0.029, $T_{eff}$ = 50%). The graph depicts the change in slope between model 1, which plots the slope term for the interaction between ΔmiR-195 and ΔComp, and model 2, which plots the slope term for the interaction between ΔmiR-195 and ΔComp when Δheptanoate was included in the regression model. **(D)** There was a positive interaction between ΔmiR-30d and Δadenosine, a negative interaction between Δadenosine and ΔComp, and a negative interaction between ΔmiR-30d and ΔComp. When Δadenosine was added to the regression model, the relationship between ΔmiR-30d and ΔComp no longer existed; therefore, adenosine significantly mediated the relationship (Sobel *p*-value = 0.001, $T_{eff}$ = 60%). The graph depicts the change in slope between model 1, which plots the slope term for the interaction between ΔmiR-30d and ΔComp, and model 2, which plots the slope term for the interaction between ΔmiR-30d and ΔComp when Δadenosine was included in the regression model. **(E)** There was a positive interaction between ΔmiR-30d and Δadenosine, a negative interaction between Δadenosine and ΔRT, and a negative interaction between ΔmiR-30d and ΔRT. When Δadenosine was added to the regression model, the relationship between ΔmiR-30d and ΔRT no longer existed; therefore, adenosine significantly mediated the relationship (Sobel *p*-value = 0.001, $T_{eff}$ = 58%). The graph depicts the change in slope between model

1, which plots the slope term for the interaction between ΔmiR-30d and ΔRT, and model 2, which plots the slope term for the interaction between ΔmiR-30d and ΔRT when Δadenosine was included in the regression model. **(F)** There was a negative interaction between ΔmiR-151-5p and Δ8-HOA, a positive interaction between Δ8-HOA and ΔComp, and a negative interaction between ΔmiR-151-5p and ΔComp. When Δ8-HOA was added to the regression model, the relationship between ΔmiR-151-5p and ΔComp no longer existed; therefore, 8-HOA significantly mediated the relationship (Sobel $p$-value = 0.007, $T_{eff}$ = 55%). The graph depicts the change in slope between model 1, which plots the slope term for the interaction between ΔmiR-151-5p and ΔComp, and model 2, which plots the slope term for the interaction between ΔmiR-1951-5p and ΔComp when Δ8-HOA was included in the regression model. **(G)** There was a negative interaction between ΔmiR-92a and Δsuberate, a positive interaction between Δsuberate and ΔComp, and a negative interaction between ΔmiR-92a and ΔComp. When Δsuberate was added to the regression model, the relationship between ΔmiR-92a and ΔComp no longer existed; therefore, suberate significantly mediated the relationship (Sobel $p$-value = 0.004, $T_{eff}$ = 68%). The graph depicts the change in slope between model 1, which plots the slope term for the interaction between ΔmiR-92a and ΔComp, and model 2, which plots the slope term for the interaction between ΔmiR-92a and ΔComp when Δsuberate was included in the regression model. **(H)** There was a negative interaction between ΔmiR-92a and Δheptanoate, a positive interaction between Δheptanoate and ΔComp, and a negative interaction between ΔmiR-92a and ΔComp. When Δheptanoate was added to the regression model, the relationship between ΔmiR-92a and ΔComp no longer existed; therefore, heptanoate significantly mediated the relationship (Sobel $p$-value = 0.006, $T_{eff}$ = 50%). The graph depicts the change in slope between model 1, which plots the slope term for the interaction between ΔmiR-92a and ΔComp, and model 2, which plots the slope term for the interaction between ΔmiR-92a and ΔComp when Δheptanoate was included in the regression model.

In total, there were 14 significant hypothesized mediations (Sobel $p$-value < 0.05, $T_{eff}$ > 50%) comprised of seven metabolites (2-HG, 8-HOA, UND, sebacate, suberate, heptanoate, adenosine), six miRNAs (miR-20a, miR-505, miR-92a, miR-151-5p, miR-195, and miR-30d), and three VR scores (Comp, Bal, and RT). Of the seven metabolites, six were FAs. Five of the six FAs (2-HG, 8-HOA, UND, sebacate, and suberate) significantly increased from pre- to postseason, while heptanoate and adenosine, a nucleoside, significantly decreased (Figure 6). Interestingly, these five measures - Comp, 8-HOA, miR-505, miR-92a, and miR-151-5p - were observed in both pre- and across-season mediations.

**Figure 6: Mediation-related metabolite changes across-season**

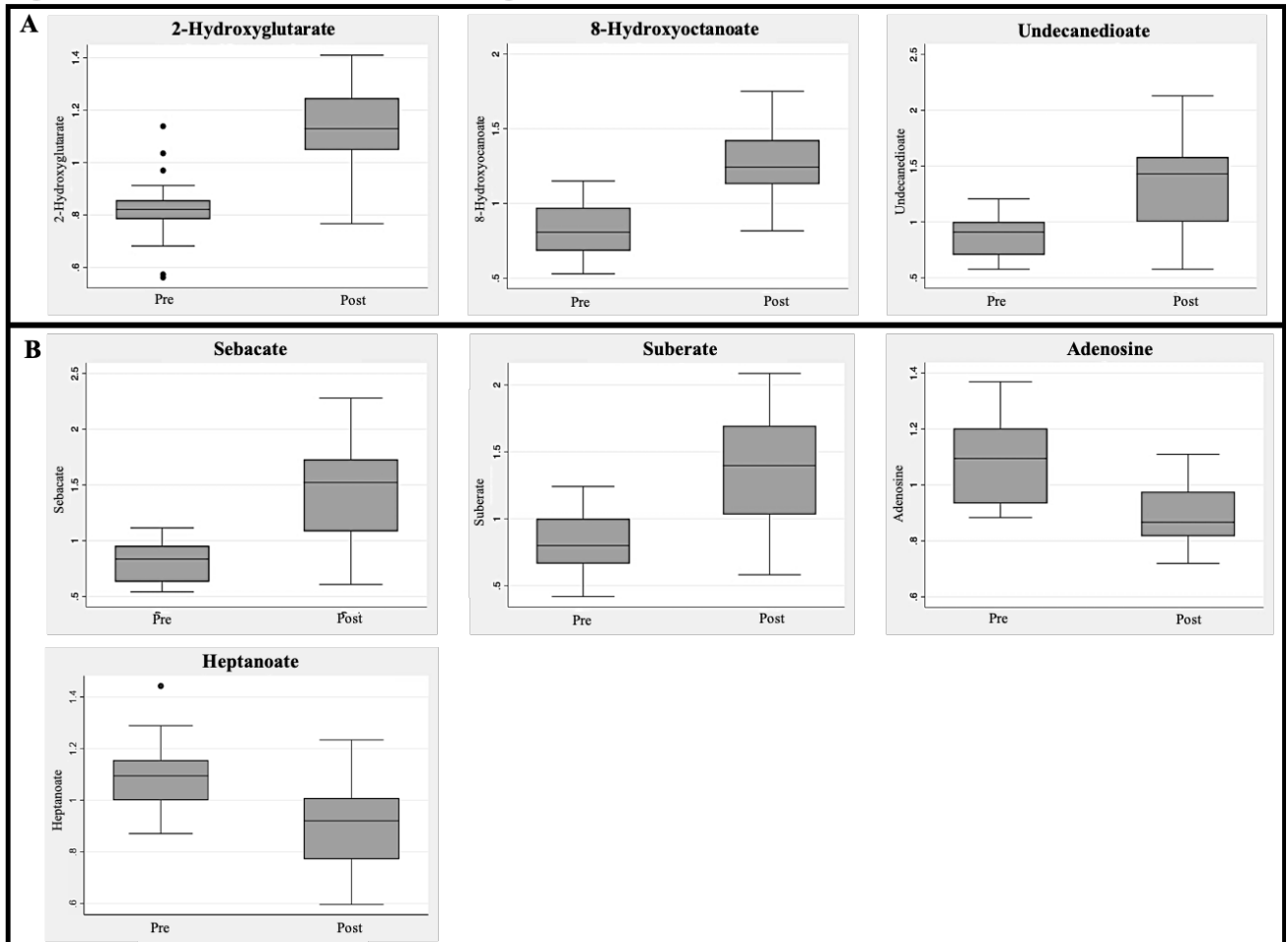

**Figure 6:** Box plots of mediation-related metabolites across season. **(A)** Metabolites associated with significant preseason mediations. 2-HG, 8-HOA, and UND significantly increased ($q$-value < 0.05) from pre- to post-season. **(B)** Metabolites associated with significant across-season mediations. Sebacate and suberate significantly increased ($q$-value < 0.05) while adenosine and heptanoate decreased from pre- to post-season.

**Summary of HAE interactions**

Two methods were used to assess HAE interactions: 1) extrapolation of subject-specific HAEs to predict preseason VR scores, miRNA levels, and metabolite levels (note: this analysis is purely exploratory and should only be considered as setting up hypotheses for future analyses), 2) assessment of how HAEs may impact across-season changes in VR scores, miRNA, and metabolites.

At preseason, there were six significant interactions between HAEs and miRNAs (5/6 related to HAEs above 80G) and two interactions between HAEs and metabolites (2/2 related to HAEs above 25G) (Table 4). Across-season, there were two significant interactions between HAEs and miRNAs (both positive; $+\beta$), as well as four interactions between HAEs and metabolites (all positive; $+\beta$) (Table 4). Interactions with miRNA were related to normalized HAEs (i.e. $aHAE_{25G}$ and $aHAE_{80G}$) while interactions with metabolites (sebacate and suberate) were related to HAEs exceeding 25G ($cHAE_{25G}$ and $aHAE_{25G}$). Significant HAE regressions were plotted in Figure 7 and their interaction terms with significant across-season mediations can be seen in Figure 2A-I.

**Table 4: Significant HAE interactions**

| Session | DV | IV (HAE) | $R^2_{adj}$ | Std. β | $p$-value | Cook's outliers |
|---|---|---|---|---|---|---|
| Pre | 505 | $cHAE_{80G}$ | 0.354 | -0.623 | 0.003 | 1/21 |
| Pre | 505 | $cHAE_{25G}$ | 0.192 | -0.535 | 0.035 | 2/21 |
| Pre | 505 | $aHAE_{80G}$ | 0.255 | -0.850 | 0.016 | 2/21 |
| Pre | 92a | $cHAE_{80G}$ | 0.382 | -0.644 | 0.002 | 2/22 |
| Pre | 92a | $aHAE_{80G}$ | 0.233 | -0.518 | 0.021 | 3/22 |
| Pre | 93p | $cHAE_{80G}$ | 0.266 | -0.550 | 0.010 | 1/22 |
| Pre | undecanedioate | $cHAE_{25G}$ | 0.144 | 0.430 | 0.046 | 1/23 |
| Pre | undecanedioate | $aHAE_{25G}$ | 0.189 | 0.477 | 0.025 | 1/23 |
| Post-Pre | Δ505 | $aHAE_{25G}$ | 0.191 | 0.486 | 0.035 | 1/20 |
| Post-Pre | Δ92a | $aHAE_{80G}$ | 0.177 | 0.475 | 0.046 | 2/20 |
| Post-Pre | Δsebacate | $cHAE_{25G}$ | 0.279 | 0.559 | 0.007 | 1/23 |
| Post-Pre | Δsebacate | $aHAE_{25G}$ | 0.190 | 0.480 | 0.028 | 2/23 |
| Post-Pre | Δsuberate | $cHAE_{25G}$ | 0.174 | 0.464 | 0.034 | 2/23 |
| Post-Pre | Δsuberate | $aHAE_{25G}$ | 0.178 | 0.468 | 0.032 | 2/23 |

**Table 4:** Across-season interactions with head acceleration events (HAEs). Four postseason HAE metrics ($cHAE_{25G}$, $cHAE_{80G}$, $aHAE_{25G}$, and $aHAE_{80G}$) were regressed against mediation-related VR scores, miRNA levels, and metabolite levels. Cook's outliers were removed prior to analyses. Results were presented if $p$-value < 0.05. Adjusted $R^2$ ($R^2_{adj.}$) and standardized beta coefficients (Std. β) were reported for each significant regression.

**Figure 7: Significant across-season HAE-related linear regression plots**

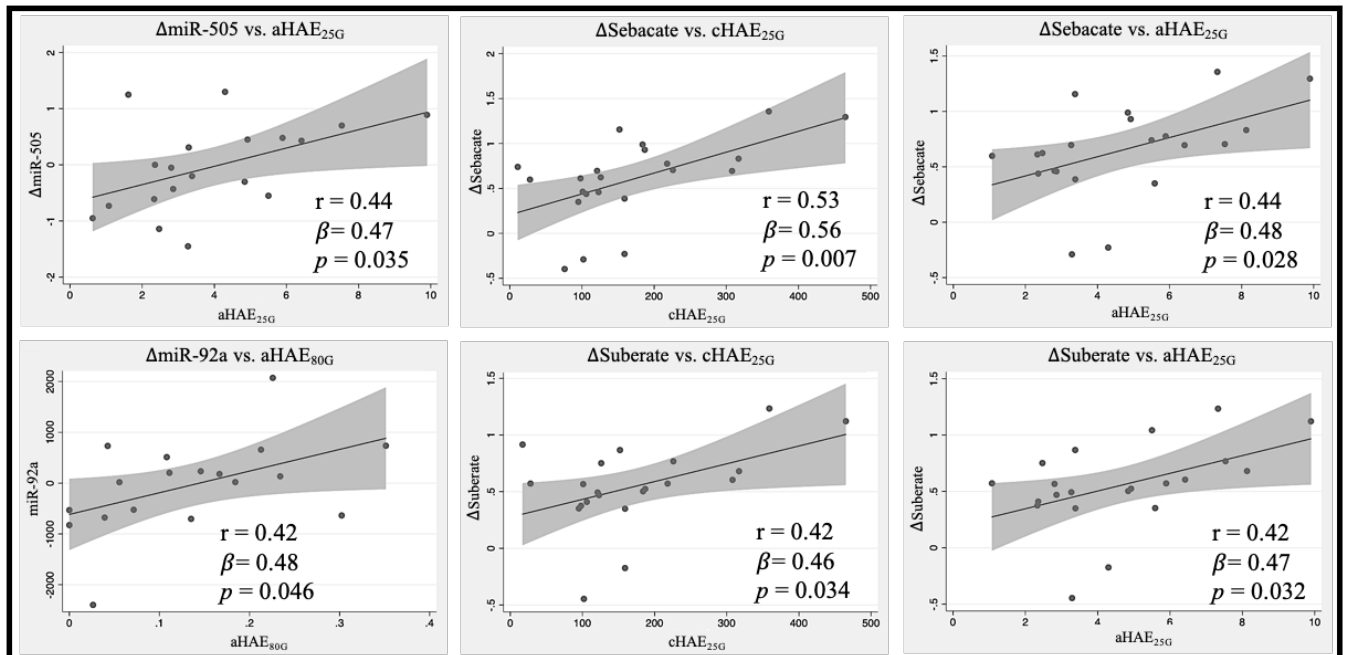

**Figure 7:** HAE-related across-season regression plots. Significant regressions from Supplemental Table 2 were graphed. Standardized beta coefficients (β; slope), *p*-values, and correlation coefficients (r) were reported for each plot. Relationships between HAEs and miRNA, or metabolite, were all positive (+β).

DISCUSSION

This study tested the hypothesis that contact athletes, who experience repetitive HAEs, would exhibit three-way associations between inflammatory-related miRNAs, metabolomic compounds, and behaviors implicated by Luria with mTBI [57, 59, 60]. We focused on metabolomic measures that were both abnormal across a season of play and (i) FAs and/or compounds involved with energy metabolism, (ii) compounds involved with stress/inflammatory responses, or (iii) exogenous compounds related to consumption. We expected these three-way associations to show statistical mediation which was consistent (i.e., highly specific) in terms of the category of independent variable (IV), mediator variable (M), and dependent variable (DV), and that computational behavior measures would always be the DV. This hypothesis was confirmed with five general findings. <u>First</u>, across-season metabolomic analysis identified 40 compounds that were significantly different across-season for three categories of focus [i.e., (i) – (iii) above], including a broad array of xanthines that were increased at the end of the season. Second, 14 mediation relationships were observed, six with preseason data, and eight with across-season data. In the all cases, except one, miRNAs were the IVs, metabolomic compounds were the mediators, and Comp or Bal behavior were the DVs. There was only one case where miRNA was the mediator of these relationships; however, the hypothesis-directed mediation was also significant indicating a mixed mediation model. In the majority of cases, biochemical processes carried the relationship between inflammatory miRNA and Luria behavior. Third, three of the miRNAs showed relationships both preseason and across-season (miR-505, miR-92a, and miR-151-5p), and 12 of the 14 mediations involved a comprehensive measure of Luria behavior (Comp). For one metabolite, 8-HOA, we observed that it significantly mediated miRNA-behavior relationships both preseason and across-season. Fourth, all seven of the metabolomic compounds mediating relationships between inflammatory miRNA and Luria behavior were those involved with some aspect of energy metabolism. Specifically, five of the seven are involved in fatty acid oxidation, a sixth compound is involved with diminishing mitochondrial respiration and respiration-coupled ATP production in cancer cells (i.e., 2-HG), and a seventh is a core structural unit of ATP (i.e., adenosine). It should be noted that altogether, these findings point to a shift mitochondria metabolism, away from mitochondria function, consistent with other illnesses classified as mitochondrial disorders [84–90].

Metabolite changes indicative of mitochondrial dysfunction and altered metabolism

Metabolites of the TCA cycle consistently decreased across-season while monohydroxy and dicarboxylic FAs increased. This observation points to dysfunctional energy metabolism, specifically related to β-oxidation. In functional mitochondria, β-oxidation breaks down FAs into acetyl-CoA which is then fed into the TCA cycle for further cellular respiration. However, several studies indicate that mitochondria are dysfunctional following mild to severe TBI [29, 31, 32, 35, 36, 91–94]. Because mitochondria are the predominate site of energy production, their dysfunction creates a state of energetic crisis and metabolism thus shifts onto other organelles. Peroxisomes are small, dynamic organelles that oxidize FAs using α-, β-, and ω-oxidative processes. ω-oxidation transforms very long, long, medium, and short-chain FAs into dicarboxylic FAs which are then β-oxidized into acetyl-CoA. However, the reported data demonstrate an accumulation of medium-chain monohydroxy (7-HOA, 8-HOA) and dicarboxylic (suberate, sebacate, UND) FAs and a decrease in TCA metabolites, suggesting that the carboxylic FAs cannot be further oxidized into acetyl-CoA for normal TCA cycle respiration (Figure 8). This observation suggests an impairment in β-oxidative processes, a resulting build-up of monohydroxy and dicarboxylic FAs, incomplete oxidation of these FAs into acetyl-CoA, and a resulting decrease in TCA-related metabolites leading to increased metabolic demand. Intriguingly, increased serum levels of 7-HOA, 8-HOA, suberate, and sebacate have been observed in patients with medium-chain acyl-CoA dehydrogenase deficiency (MCADD) – a genetic mitochondrial disorder caused by a mutation in the ACADM gene [95–97]. Interestingly, another feature of this disorder is increased serum levels of acyl-carnitine derivatives with 6, 8, and 10 carbons. Indeed, these metabolites were increased in our cohort; however, most $p$-values were only trending ($p$-value > 0.05) and a larger cohort must be studied to further validate this observation (Table 5). The last FA observed in across-season mediation relationships was heptanoate. Heptanoate is a monohydroxy carboxylic fatty acid that decreased across the season of football – opposite of what was observed for the other monohydroxy and dicarboxylic fatty acids. Interestingly, increased levels of heptanoate, via metabolism of triheptanoin, has been marked as beneficial in the treatment of $\beta$-oxidation related metabolic disorders [98–100]. Therefore, an observed decrease in heptanoate could be detrimental.

In relation to brain injury, one study observed increased neuronal cell loss with the accumulation of free FAs in rat models of TBI [101]. Taken together, the observed increase in medium-chain monohydroxy and dicarboxylic FAs, decrease in heptanoate, and concurrent decrease in TCA metabolites suggest an impairment in $\beta$-oxidation stemming from injury-induced mitochondrial dysfunction, which has been reported in several TBI-related studies [34, 102].

**Figure 8: Shift in metabolism**

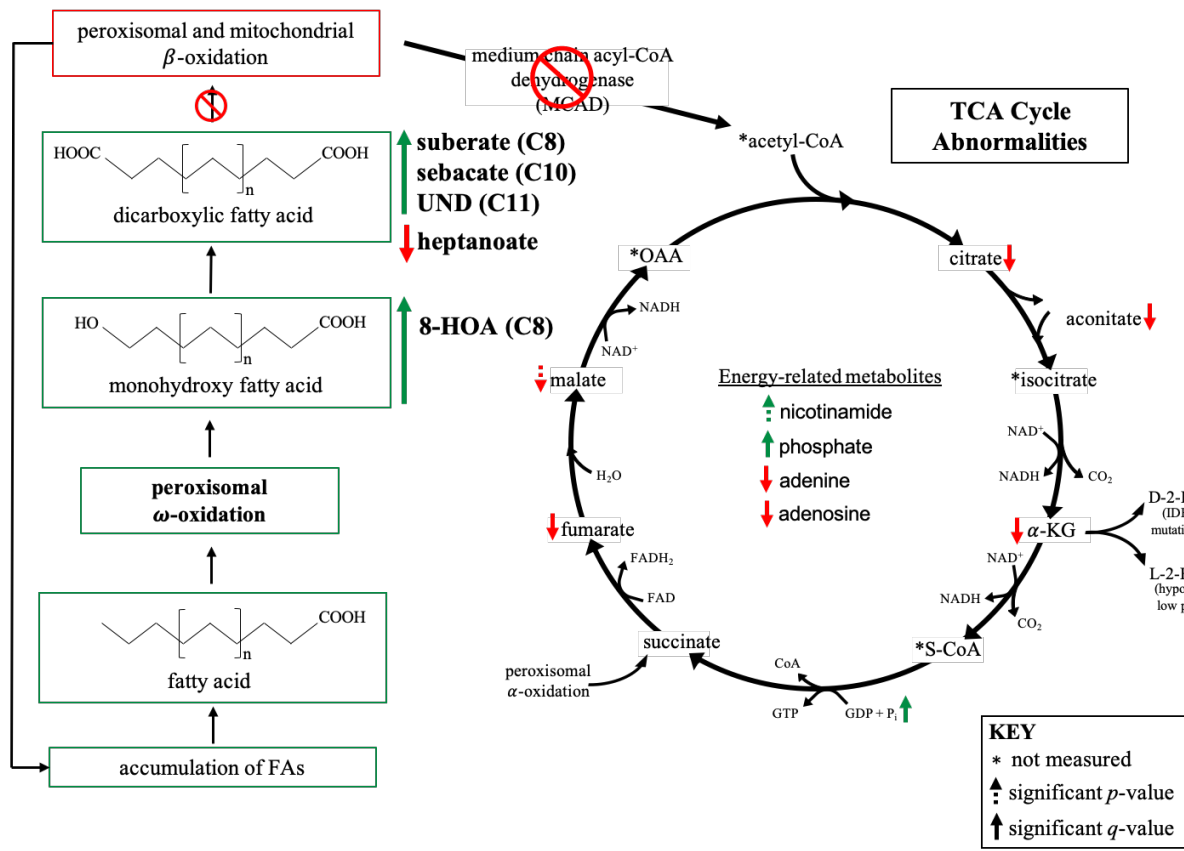

**Figure 8:** Synopsis figure summarizing the observed shift in metabolism and its relationship to inflammatory miRNAs and complex behavior. There were significant increases in medium-chain monohydroxy and dicarboxylic fatty acids (FAs) (suberate, sebacate, UND, and 8-HOA) from pre- to post-season. Increases in these FAs have been associated with genetic disorders related to impaired beta-oxidation. Impairment of this critical process can result in an accumulation of medium-chain FAs which cannot be further oxidized into smaller, functional, metabolites, such as acetyl-CoA. Acetyl-CoA is a critical starting-point metabolite for the TCA cycle, a major source of energy-rich molecules that are fed

into further energy-producing processes (e.g., electron chain transport system). Here, TCA-related metabolites (citrate, aconitate, $\alpha$-KG, fumarate, and malate) were all observed to decrease, suggesting a problem with the initial step of the cycle (i.e. lack of acetyl-CoA). Additionally, there were alterations in energy-rich molecules such as adenine, adenosine, nicotinamide, and phosphate, suggesting a state of energy imbalance. Lastly, 2-HG, a known oncometabolite, was observed to increase. Regardless of its role as an oncometabolite, its increase suggests a state of oxidative stress. Taken together, it is suggested that there are dysfunctional beta-oxidative processes in this cohort of collegiate football players leading to subsequent issues with energy production.

**Table 5: Wilcoxon signed-rank test for acyl-carnitine derivatives**

| Metabolite | $p$-value | $q$-value | z-score | outliers |
|---|---|---|---|---|
| octanoylcarnitine (C8) | 0.1283 | 0.1283 | -1.521 | 0/23 |
| decanoylcarnitine (C10) | **0.0335** | 0.1005 | -2.127 | 1/23 |
| cis-4-decenoylcarnitine (C10) | **0.0150** | 0.0652 | -2.433 | 0/23 |
| adipoylcarnitine (C6) | **0.0163** | 0.0652 | -2.403 | 0/23 |
| suberoylcarnitine (C8) | 0.0795 | 0.1283 | -1.753 | 1/23 |

**Table 5:** Changes in acyl-carnitine derivatives. The number of carbons in each metabolite is noted in parentheses. Cook's outliers were removed prior to regression analysis and the number of outliers removed in each regression is listed as the ratio of outliers to the total n. $p$-Values were reported at a significance level of 0.05 and a Benjamini-Hochberg FDR correction was applied to obtain $q$-values and. Negative z-scores confer an increase from pre- to post-season.

Additionally, 2-HG, a known oncometabolite, increased from pre- to postseason. 2-HG is dicarboxylic fatty acid generated from $\alpha$-ketoglutarate (i.e., oxoglutaric acid), a component of the TCA cycle, via the mutated IDH gene [103, 104]. 2-HG can exist in two enantiomeric forms (L and D) and both have been implicated in tumor suppressor inactivation and oncogenic activation [105–107]. In addition to its oncogenic function, 2-HG is an established marker of oxidative stress and has been associated with poorer outcomes following brain injury [108, 109]. Interestingly, L-2-HG structurally resembles glutamate and thus may affect neurotransmission [110]. Here, an increase in 2-HG is likely linked to oxidative stress resulting from repetitive HAEs and may even alter neurotransmission and subsequent network connectivity [111–115].

In this study, we observed adenosine to be decreased at postseason. Adenosine is a potent vasodilator that increases regional cerebral blood flow (rCBF), thereby decreasing energy demands. It is also implicated as a robust neuroprotector and has been reported to suppress ATP release [116]. Further, dysregulation of adenosine levels has been reported following TBI and is linked with numerous TBI-related pathologies and comorbidities [117]. In particular, severe TBI in humans was associated with increased adenosine levels, which was hypothesized to play a neuroprotective role and act to uncouple rCBF from oxidative metabolism [118]. In rats, adenosine agonists administered 10 minutes post-TBI reduced injury severity following fluid-percussion TBI [119]. However, adenosine has also been reported to decrease 24 hours following TBI in rats [116], and decreases in extracellular adenosine have been associated with a *loss* of neuroprotection during states of hypoxia and ischemia [120]. These, and other, findings suggest an interplay between protective effects of adenosine receptors A1 and $A_{2A}$, and further research is required to elucidate time course effect of adenosine across the full spectrum of TBI. Taken together, our findings suggest that exposure to repetitive subconcussive HAEs may trigger a detrimental decrease in adenosine and subsequent neuroprotective dysfunction.

Another energy-related metabolite, phosphate, was found to increase in the absence of mediation relationships. This observation may be associated with reduced ATP synthesis, ATP depletion, and/or ATP hydrolysis. In cases of ischemia, mitochondria hydrolyzed ATP more frequently [121] and ATP synthesis was impaired in animals of brain injury [122]. Impaired ATP synthesis may be related to depleted energy stores and a subsequent lack of energy sources. Contrary to these observations, serum phosphate levels were shown to decrease in patients with *severe* TBI [123, 124], yet football athletes experience repetitive *subconcussive* HAEs. More research is required to elucidate the role of phosphate in the full spectrum of TBI.

Lastly, all xanthine metabolites were increased at postseason suggesting either 1) an increase in caffeine consumption [35, 125–128], 2) activation of xanthine oxidases [129–132], or

3) a combination of both [29]. Caffeine is an exogenous compound shown to have anti-inflammatory and neuroprotective effects, especially in the context of neurodegenerative disease and TBI [133]. In human studies, caffeine levels were correlated with better outcomes in patients with severe TBI [134]; this finding may be related to caffeine-induced neuroprotection via blockage of the adenosine receptor, $A_{2A}$ [135, 136]. Additionally, caffeine consumption reduced cognitive decline in older women [137], has been suggested to reduce the likelihood of developing Alzheimer's disease [138], and has been linked to improving psychiatric disorder symptoms [139]. Moreover, numerous animal studies have demonstrated neuroprotective effects of caffeine following TBI [140–142]. However, the timing of caffeine administration may be critical, as was seen in a study where caffeine administered prior to TBI was beneficial while caffeine administered post-TBI was detrimental [143]. Increased caffeine consumption in this cohort could be the result of 1) school-related stress and lack of sleep, 2) compensation for HAE-induced brain damage and metabolic dysfunction, or 3) a combination of both. Given the current data, it is not possible to identify the exact reason for the increased levels of xanthines and further research is required. It should be noted though that previous studies have reported caffeine consumption in contact athletes with the main motive being regaining energy lost during exercise [144, 145].

Metabolite changes mediate the relationship between inflammatory miRNA and Luria behavior

Previous studies have reported increased levels of inflammatory miRNAs prior to contact play and therefore, no changes observed across the season [16, 80]. Additionally, behavioral changes alone are rarely observed and difficult to replicate in studies of contact athletes without diagnosed concussion [13, 146–149]. However, research points to significant neuroimaging (i.e. fMRI, rs-fMRI, DTI, MRS, CVR, rCBF) and biochemical changes in this population across a season of sports participation [3–5, 14, 81, 150–152]; however, behavior is not commonly shown to correlate with these changes [3, 13, 149, 153]. Therefore, it was hypothesized that metabolomic changes would mediate the relationship between elevated miRNA and potentially subtle changes in behavior that could be quantified using virtual reality technology.

At preseason, there were six significant hypothesis-directed mediations. The metabolites mediating the relationship between chronically elevated miRNA and behavior were monohydroxy (8-HOA) and dicarboxylic FAs (2-HG and UND). According to the literature, these types of FAs were elevated in patients with defective $\beta$-oxidative metabolism [85, 87, 90, 154–156]. The accumulation of these specific FAs suggests a state of chronic metabolic dysregulation related to ineffective $\beta$-oxidation in mitochondria and peroxisomes. Ultimately, this dysfunction may result in an energy crisis as FAs cannot be fully oxidized into acetyl-CoA for normal cellular respiration. In fact, decreased levels of acetyl-CoA, as well as increased phosphate and decreased ATP, have been observed in rats exposed to repetitive mTBI [29, 32]. The fact that these FAs are mediating the miRNA-behavior relationship suggests that dysfunctional $\beta$-oxidation may underlie subtle changes in behavior. Moreover, the fact that these relationships were observed *prior to* contact practice and play suggests chronic metabolic dysregulation and neuroinflammation that may persist long after season commencement and affect neurological function (Figure 9).

**Figure 9: Synopsis of observations**

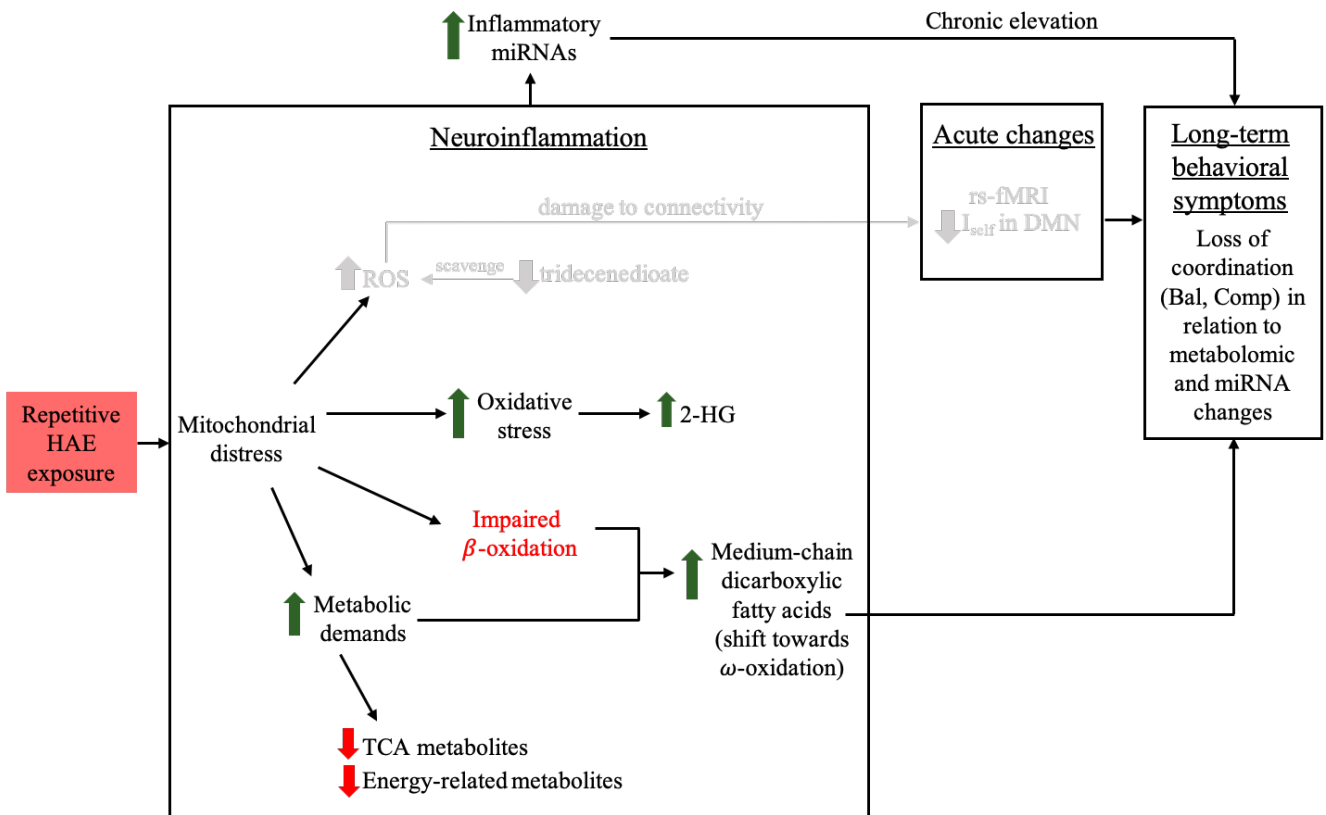

**Figure 9:** Repetitive exposure to subconcussive head impacts (or HAEs) has been shown to produce significant changes in brain homeostasis, such as increased neuroinflammation [7, 15, 157–163]. Here, we observed changes indicative of mitochondrial distress as evidenced from both the accumulation of medium-chain FAs and the subsequent decreases in TCA-related metabolites. Mitochondrial dysfunction can lead to numerous physiological disturbances, some of which were observed in the present study: 1) oxidative stress (i.e., increased levels of 2-HG), 2) impairment in beta-oxidative processes (i.e., increased levels of medium-chain FAs), and 3) increased metabolic demands (i.e., decreased TCA and energy-related metabolites). Together, these neuroinflammatory processes may be related to the observed elevation in inflammatory-related miRNA molecules (specifically miR-20a, miR-505, miR-151-5p, miR-30d, miR-92a, and miR-195). In fact, these miRNAs were significantly correlated with the metabolites shown in 3A. In addition, the metabolites were shown to mediate the relationship between elevated miRNA levels and complex Luria behavior (i.e., computation virtual reality tasks). Together, the mediating effect of these energy-related metabolites are critical when defining the relationship between elevated miRNAs and behavioral outcomes. This complex relationship may explain why obvious behavioral changes in subconcussed athletes have not been routinely observed, but how repetitive, long-term exposure to HAEs, chronic elevation of inflammatory-miRNAs, and acute, but deleterious changes in energy metabolites could result in behavioral disturbances later in life.

Across-season, there were eight significant hypothesis-directed mediations. Again, the majority of mediating metabolites were FAs that were increased across the season and may be related to dysfunctional mitochondrial β-oxidation. However, heptanoate and adenosine were also observed in mediation relationships, as discussed in the previous section. The observation of significant mediations using change measures across the season indicate that chronically elevated miRNAs, as observed in [16], and dynamic metabolic disruption appear to impact neurological function. Interestingly, three miRNAs (miR-505, miR-92a, and miR-151-5p) and one metabolite (8-HOA) were involved in both pre- and across-season mediations and 12 of the 14 mediations involved the comprehensive measure of Luria-based behavior.

Together, all seven of the metabolomic compounds mediating relationships between inflammatory miRNAs and Luria behavior were involved in some aspect of energy metabolism. Additionally, other metabolites that were not observed in mediation relationships, but are involved in energy metabolism, were observed to change (e.g., citrate, phosphate; Figure 3A). These observations strongly argue for a state of metabolic dysregulation, potentially related to dysfunctional mitochondria and subsequent dysfunction in cellular respiration (i.e., TCA cycle and downstream energy-producing processes).

<u>HAEs were associated with changes in metabolites and miRNA</u>

The observed relationships between miRNAs, metabolites, and behavior were related to HAE exposures throughout the football season (Table 4). Specifically, across-season measures of miRNAs 505 and 92a were significantly related to cumulative and average HAEs exceeding 25G and 80G. Additionally, across-season measures of the metabolites sebacate and suberate were associated with HAEs exceeding 25G. All across-season relationships were positive, indicating that increased HAE exposure was related to larger increases in neuroinflammatory miRNA levels and increased metabolomic measures (sebacate and suberate). Preseason results indicate an opposite trend where higher extrapolated HAEs were associated with lower levels of miRNA. This is interesting, as it suggests that lower levels of inflammatory-related miRNA may be associated with more HAE exposures from the previous season of play. However, these results are purely explorative as previous HAEs were not recorded and should only be considered as setting up hypotheses for future investigation.

Previous research, specifically in the neuroimaging field, has demonstrated the role HAEs play in neurophysiological changes such as altered connectivity (rs-fMRI and fMRI), neurochemical alterations (MRS), altered cerebral blood flow (perfusion imaging), and axonal injury (DTI) [4, 7, 168, 13, 15, 80, 151, 164–167]. Here, we present blood biomarkers (metabolites and miRNAs) that also correspond to HAE exposure. Interestingly, no behaviors were associated with HAEs. This absence of association may be due to inherent limitations of the sensors, a small sample size, and/or the fact that competition-related HAEs were not recorded.

Other limitations of this study include the time course of sample collection and a lack of age-matched, non-contact athlete controls. Blood collections and VR tests were only conducted twice – once prior to contact practice and once following the competition season. To observe more transient changes in these metrics, it would be beneficial to collect data at more time points during and after the season. Additionally, this study lacked age- and gender-matched, non-contact athlete controls. Previous reports of miRNA elevations in football athletes utilized healthy, non-age- and gender-matched controls [16]. Future studies should incorporate age- and gender-match controls that also participate in competitive, non-contact athletics. Lastly, while the sample size was small, a rigorous permutation-based statistical approach was applied to account for this limitation while controlling for false positives and false negatives.

CONCLUSION

The presented study evaluated the relationships between energy-related metabolites, inflammatory-related miRNAs, and Luria-based computational behavior at the preseason and across the season. Specifically, metabolites were hypothesized to mediate the relationship between elevated miRNAs and VR task performance. The majority of mediation findings involved a fatty acid (2-HG, 8-HOA, UND, sebacate, suberate, and heptanoate) and in addition, TCA metabolites were found to be significantly decreased at postseason relative to preseason. Lastly, HAEs were associated with metabolomic measures and miRNA levels across-season. Together, these observations suggest a state of chronic HAE-induced neuroinflammation (as evidence by elevated miRNA, [16]) and mitochondrial dysfunction (as observed by altered metabolomic measures) that together produce subtle changes in neurological function (as observed by behavioral relationships) (Figure 9). Additionally, these complex, multi-scale observations were discovered using an integration of permutation-based statistics with mediation analysis; this method may provide a technique for other human-focused studies of functional brain illness in lieu of, or in parallel with, animal studies. These findings provide preliminary evidence that 14 metabolites, related to potential mitochondrial distress, have complex relationships to neuroinflammation and subsequent neurological alterations. Therefore, this set of metabolites 1) supports the hypothesis of mitochondrial and metabolic dysfunction in subconcussed contact-sport athletes and 2) may serve as a promising set of biomarkers for HAE-related neurological change that does not produce obvious behavioral deficits.

CONFLICT(S) OF INTEREST

The authors have no conflicts of interest to declare.